\documentclass[11pt]{article}

\usepackage{amsmath,amsfonts,amssymb,natbib,hyperref,multirow,rotating,array}
\usepackage{enumerate,amsthm,subcaption,url,xcolor}
\newcolumntype{H}{>{\setbox0=\hbox\bgroup}c<{\egroup}@{}}
\setcitestyle{aysep={}}
\hypersetup{
    colorlinks=true,
    linkcolor=blue,
    citecolor=blue,
    filecolor=blue,
    urlcolor=blue,
}
\newcommand\crule[3][black]{\textcolor{#1}{\rule{#2}{#3}}}

\usepackage{graphicx}

\newcommand{\mvec}{\text{\bf m}}

\newcommand{\xvec}{\text{\bf x}}
\newcommand{\zvec}{\text{\bf z}}

\newcommand{\wvec}{\text{\bf w}}
\newcommand{\Zvec}{\text{\bf Z}}

\newcommand{\Xvec}{\text{\bf X}}

\newcommand{\Yvec}{\text{\bf Y}}

\newcommand{\yvec}{\text{\bf y}}

\newcommand{\sigmat}{\mbox{\boldmath $\Sigma$}}
\newcommand{\muvec}{\mbox{\boldmath $\mu$}}

\newcommand{\alphavec}{\mbox{\boldmath $\alpha$}}

\newcommand{\betavec}{\mbox{\boldmath $\beta$}}
\newcommand{\Gammavec}{\mbox{\boldmath $\Gamma$}}

\newcommand{\argmax}{\operatornamewithlimits{argmax}}
\newcommand{\argmin}{\operatornamewithlimits{argmin}}

\newtheorem{theorem}{Theorem}
\newtheorem{lemma}{Lemma}
\newtheorem{remark}{Remark}
\newtheorem{defn}{Definition}

\begin{document}

\def\spacingset#1{\renewcommand{\baselinestretch}%
{#1}\small\normalsize} \spacingset{1}


\begin{center}
{\Large\sc On Perfect Clustering of High Dimension,\\
Low Sample Size Data}
\vspace{0.05in}

\sc Soham Sarkar\footnote{Email:sohamsarkar1991@gmail.com} and Anil K. Ghosh\footnote{Email:akghosh@isical.ac.in}
\vspace{0.02in}

{\small\it Theoretical Statistics and Mathematics Unit\\ Indian Statistical Institute\\
    203, B. T. Road, Kolkata 700108}
\end{center}

\begin{abstract}
Popular clustering algorithms based on usual distance functions (e.g., Euclidean distance) often suffer in high dimension, low sample size (HDLSS) situations, where concentration of pairwise distances has adverse effects on their performance. In this article, we use a dissimilarity measure based on the data cloud, called MADD, which takes care of this problem. MADD uses the distance concentration phenomenon to its advantage, and as a result, clustering algorithms based on MADD usually perform better for high dimensional data. Using theoretical and numerical results, we amply demonstrate it in this article.

We also address the problem of estimating the number of clusters. This is a very challenging problem in cluster analysis, and several algorithms have been proposed for it. We show that many of these existing algorithms have superior performance in high dimensions when MADD is used instead of the Euclidean distance. We also construct a new estimator based on penalized Dunn index and prove its consistency in the HDLSS asymptotic regime, where the sample size remains fixed and the dimension grows to infinity. Several simulated and real data sets are analyzed to demonstrate the importance of MADD for cluster analysis of high dimensional data.

\vspace{0.1in}
\noindent%
{\bf Keywords:}  Dunn index, hierarchical clustering, high dimensional consistency, k-means clustering, pairwise distances, Rand index.
\end{abstract}

\spacingset{1.45}

\vspace{0.05in}
\section{Introduction}\label{sec::intro}
\vspace{-0.05in}

Let $\xvec_1,\ldots,\xvec_n \in {\mathbb R}^d$ be a sample of $n$ unlabeled observations coming from different populations. The aim of cluster analysis is to divide this sample into several groups of `similar' observations. In practice, one uses an appropriate measure of similarity (or, dissimilarity) between a pair of observations, and a clustering algorithm is developed based on that. When all measurement variables are continuous, a popular choice for the dissimilarity index is the Euclidean distance or the squared Euclidean distance. Popular clustering algorithms like $k$-means, $k$-medoids and hierarchical clustering \citep[see][]{FHT01,DHS12} generally use dissimilarity indices based on the Euclidean distance. Spectral clustering algorithms \citep[see][]{L07} often use similarity index based on the radial basis function, which is a decreasing function of the Euclidean distance. These algorithms work
well when the sample size is sufficiently large. But, like other nonparametric methods, they often perform poorly in high dimension, low sample size (HDLSS) situations.

\begin{figure}[h!]
\centering
\includegraphics[height=2.250in,width=5.75in]{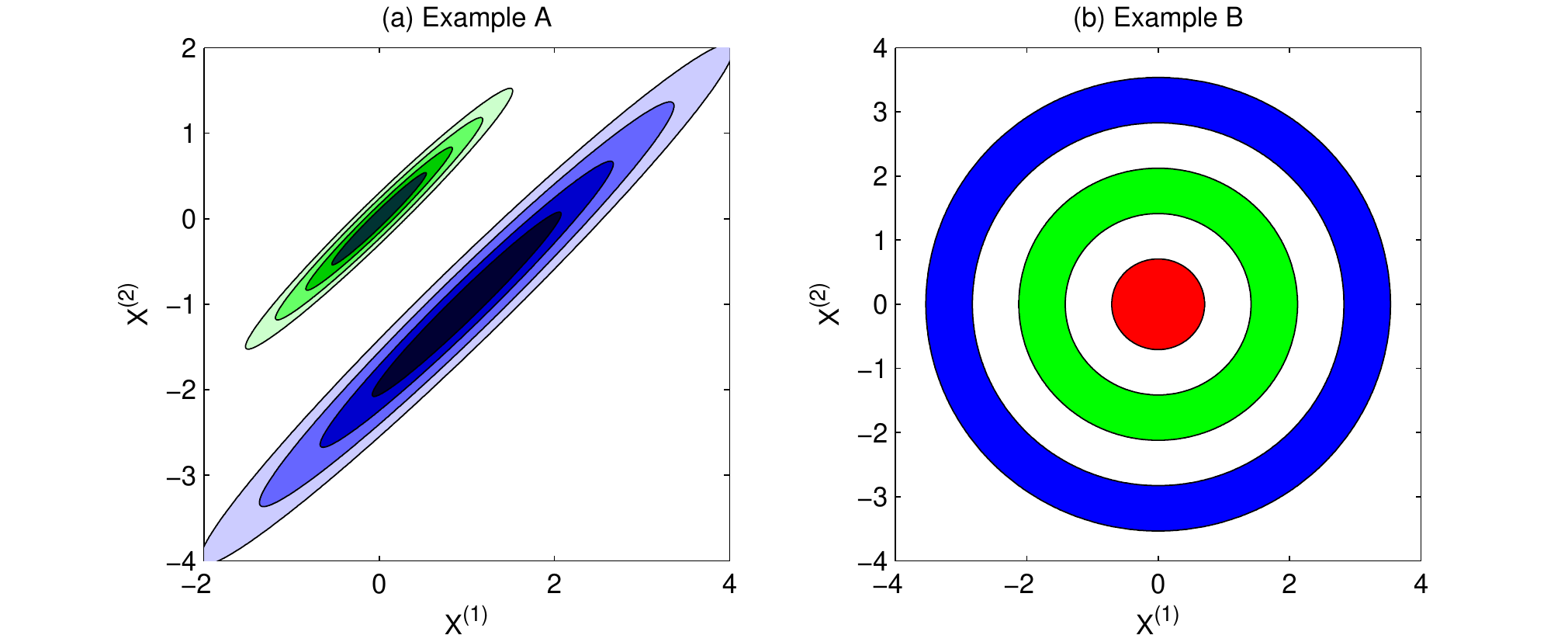}

\vspace{-0.15in}
\caption{(a) Central regions of the two normal populations in Example A  and (b) Supports of the three non-overlapping populations in Example B (for $d=2$).}
\label{fig:examplesA&B}
\vspace{-0.15in}
\end{figure}

\begin{figure}[h!]
\centering
\includegraphics[height=2.40in,width=6.0in]{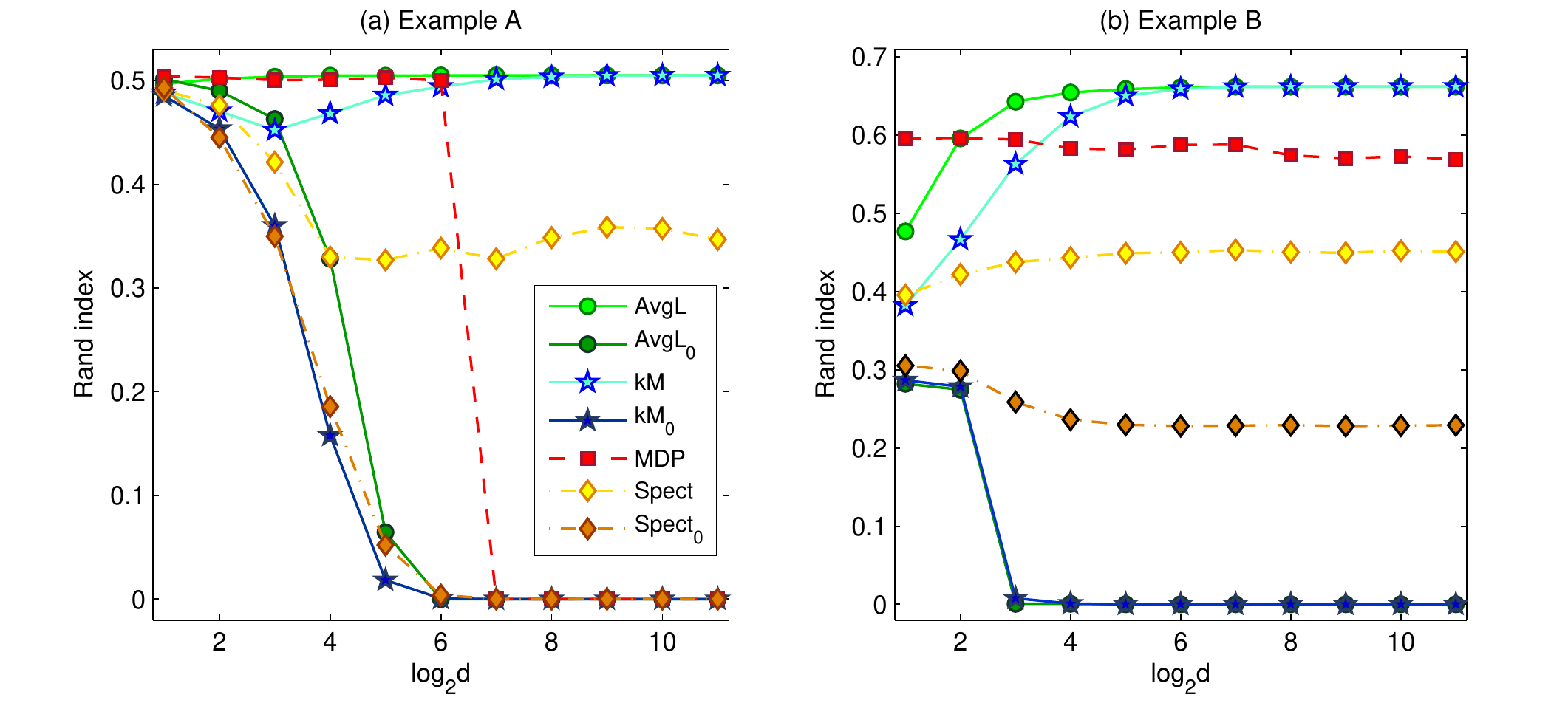}

\vspace{-0.2in}
\caption{Rand indices for different algorithms in Examples A and B.}
\label{fig:rand}
\vspace{-0.1in}
\end{figure}

To demonstrate this, we consider an example (call it Example A), with two $d$-dimensional normal distributions ${\cal N}_d({\bf 0}_d,\sigma_1^2{\sigmat}_d)$ and ${\cal N}_d(\muvec_d,\sigma_2^2{\sigmat}_d)$, where ${\bf 0}_d=(0,\ldots,0)^\top$, $\muvec_d=(1,-1,\ldots,(-1)^{d+1})^\top$, and $\sigmat_d=((\sigma_{ij}))_{d \times d}$ is a block diagonal matrix with $\sigma_{ii}=1$ for $i=1,\ldots,d$, $\sigma_{(2i-1)2i} = \sigma_{2i(2i-1)} = 0.98$ for $i=1,\ldots,\lfloor d/2\rfloor$ ($\lfloor t \rfloor$ is the largest integer $\le t$) and $\sigma_{ij}=0$ otherwise. Taking $\sigma_1^2=0.5$ and $\sigma_2^2=2$, we generated 50 observations from each distribution. Figure~\ref{fig:examplesA&B}(a) shows the central regions of these two distributions with coverage probabilities $0.25,~ 0.5,~ 0.75$ and $0.9$ when $d=2$. We used the average linkage method (AvgL) as well as the $k$-means algorithm (kM) based on Euclidean distance to estimate two clusters in the sample consisting of 100 observations.  For $i=1,\ldots,n$, let ${\cal C}(\xvec_i)$ be the actual cluster label of $\xvec_i$ and $\delta(\xvec_i)$ be the cluster label assigned to $\xvec_i$ by a clustering algorithm $\delta$. We measure the performance of $\delta$ using the Rand index \citep[see][]{R71}
\begin{equation}\label{eqn::Rand}
{\cal R}(\delta)={\binom{n}{2}}^{-1} \sum_{1 \le i < j \le n} \mathbb{I}\bigg[\mathbb{I}\{\delta(\xvec_i)=\delta(\xvec_j)\}+\mathbb{I}\{{\cal C}(\xvec_i)={\cal C}(\xvec_j)\}=1\bigg],
\vspace{-0.05in}
\end{equation}
where $\mathbb{I}\{\cdot\}$ is the indicator function. Note that $\delta$ leads to perfect clustering if  $\delta(\xvec_i)=\pi\{{\cal C}(\xvec_i)\}$ for all $i=1,\ldots,n$ and a suitable permutation $\pi$ of $\{1,\ldots,\max\{{\cal C}(\xvec_1),\ldots,{\cal C}(\xvec_n)\}\}$. In that case, we have ${\cal R}(\delta)=0$. Higher values of Rand index indicate more deviation from perfect clustering. We repeated our experiment 100 times, and the average  Rand index  of an algorithm over these 100 trials was computed for $d=2^r$, with $r=1,\ldots,11$. In this example, separation between the two populations is quite evident (see Figure~\ref{fig:examplesA&B}(a)), and it increases with the dimension.
So, for a good clustering algorithm, the Rand index is expected to shrink to $0$ as $d$ increases.  But, that was not the case for AvgL and kM. Both of them had miserable performance for all values of $d$ (see Figure~\ref{fig:rand}(a)). The spectral clustering algorithm proposed by \cite{SM00} (Spect) also had higher Rand index when a similarity measure based on radial basis function was used (see Figure~\ref{fig:rand}(a)).

We carried out another experiment with observations generated from three distributions with disjoint supports, viz., ${\cal U}_d(0,0.5)$, ${\cal U}_d(1,1.5)$ and ${\cal U}_d(2,2.5)$. Here ${\cal U}_d(a,b)$ denotes the $d$-dimensional uniform distribution over the region $\{\xvec \in \mathbb{R}^d : a\sqrt{d}\le \|\xvec\| \le b\sqrt{d}\}$. Figure \ref{fig:examplesA&B}(b) shows the supports of these three distributions for $d=2$. We generated 50 observations from each distribution, and different clustering algorithms were used to divide these 150 observations into three different clusters. In this example (call it Example B) also, all these three methods, especially AvgL and kM, had poor performance (see Figure~\ref{fig:rand}(b)).

Clustering algorithm based on maximal data pilling (MDP) distance \citep{Ahn12}, which is especially designed for high dimensional data, performed well in Example A for $d \ge 2^7$, but it performed poorly for all smaller values of $d$. In Example B, its performance was even worse. It had much higher Rand index for all values of $d$ considered here.

Failure of these popular algorithms shows the necessity to develop new methods for clustering high dimensional data. In this article, we use a new dissimilarity index, called MADD (defined in Section~\ref{sec::MADD}), for this purpose. Notice that both AvgL and kM yielded excellent results when MADD was used as the dissimilarity measure (see the curves corresponding to AvgL$_0$ and kM$_0$ in Figure~\ref{fig:rand}). In both examples, they led to perfect clustering in high dimensions. It is well known that both AvgL and kM based on Euclidean distance are not much useful for finding non-convex clusters in the data, but Example B clearly shows that their MADD versions can overcome this limitation.
Spectral clustering algorithm of \cite{SM00} also performed well when a decreasing function of MADD (defined in Section~\ref{sec::MADD}) was used as the similarity measure. While the use of MADD led to perfect clustering for large $d$ in Example A, it significantly reduced the Rand index in Example B as well (see the curves corresponding to Spect$_0$ in Figure~\ref{fig:rand}).

The reasons behind the failure of Euclidean distance based clustering and the excellence of MADD based clustering are investigated in Section~\ref{sec::MADD}. In this connection, we prove the high dimensional consistency of some clustering algorithms based on MADD. Simulation studies are also carried out to demonstrate the superiority of these MADD based algorithms. We consider the problem of estimating the number of clusters from the data in Section~\ref{sec::no-of-clusters}. This is an important problem in cluster analysis, and several methods are available for it. We observe that many of these methods perform better when MADD is used for their constructions. We investigate the high dimensional behavior of these MADD based estimation methods under appropriate regularity conditions. We also construct a new estimator based on penalized Dunn index and prove its high dimensional consistency. Empirical performances of different estimation methods are evaluated using simulation studies. Two benchmark data sets are analyzed in Section~\ref{sec::benchmark} for further comparison among different estimation methods and clustering algorithms. Finally, Section~\ref{sec::remarks} gives a brief summary of the work and ends with some related discussions. All proofs and mathematical details are given in the Appendix. 

\vspace{-0.175in}
\section{Clustering algorithms based on MADD}\label{sec::MADD}

\vspace{-0.075in}
Suppose that the whole sample ${\cal X} = \cup_{i=1}^{k_0} {\cal X}_i$ 
consists of $n$ unlabeled observations, where ${\cal X}_i$ denotes the collection of $n_i$ observations ($\sum_{i=1}^{k_0} n_i =n$) from the $i$-th ($i=1,\ldots,k_0$) population. In Example A, we had $k_0=2$ and $n_1=n_2=50$. In this example, for two observations $\Xvec=(X^{(1)},\ldots,X^{(d)})^{\top}$ and $\Yvec=(Y^{(1)},\ldots,Y^{(d)})^{\top}$ from the second population, $d^{-1}\|\Xvec-\Yvec\|^2 =d^{-1}\sum_{q=1}^{d}(X^{(q)}-Y^{(q)})^2$, being an average of an 
$m$-dependent sequence (with $m=1$) of identically distributed random variables, converges to $E(X^{(1)}-Y^{(1)})^2=2\sigma_2^2=4$ in probability. But, if $\Xvec$ comes from the first population and $\Yvec$ comes from the second population, $d^{-1}\|\Xvec-\Yvec\|^2$ converges in probability to $\sigma_1^2+\sigma_2^2+1=3.5$. Due to this concentration of pairwise distances, for large $d$, all observations in ${\cal X}_2$ had their neighbors in ${\cal X}_1$. So, clustering algorithms based on the Euclidean distance failed to put them in the same cluster. \cite{HMN05} proved the concentration of Euclidean distance assuming weak dependence among the component variables and provided an idea about the high dimensional geometry of the data cloud consisting of observations from two distributions. They also pointed out the adverse effects of distance concentration on some popular classifiers. In Example A, we observe its adverse effects on clustering algorithms. In Example B also, we have similar convergence of pairwise distances, which is shown in the following lemma.
\vspace{-0.05in}
\begin{lemma}\label{lemma::dist-convergence}
If $\Xvec \sim {\cal U}_d(a_1,b_1)$, $\Yvec \sim {\cal U}_d(a_2,b_2)$, and they are independent, then $d^{-1}\|\Xvec-\Yvec\|^2 $ converges in probability to $b_1^2 + b_2^2$ as the dimension $d$  tends to infinity.
\end{lemma}
\vspace{-0.1in}
From Lemma~\ref{lemma::dist-convergence}, it is clear that
for large values of $d$, all observations in ${\cal X}_2$ and ${\cal X}_3$ had their nearest neighbors in ${\cal X}_1$ only. So, AvgL and kM algorithms had miserable performances in this example as well. However, these two algorithms produced excellent results in Examples A and B when, instead of Euclidean distance, we used a new dissimilarity index given by $\rho(\xvec,\yvec) = (n-2)^{-1} \sum_{\zvec \in {\cal X} \setminus \{\xvec,\yvec\}} \bigl|\|\xvec - \zvec\| - \|\yvec - \zvec\|\bigr|$. Following our above discussion, one can show that in both of these examples,
$d^{-1/2} \rho(\Xvec,\Yvec)\stackrel{P}{\rightarrow} 0$ as $d \rightarrow \infty$ if and only if $\Xvec$ and $\Yvec$ come from the same population. Otherwise, it converges to a positive constant. So, clustering algorithms based on $\rho$ had better performance in high dimensions.

This type of dissimilarity index based on {\bf M}ean {\bf A}bsolute {\bf D}ifference of {\bf D}istances (called MADD) can be constructed using other distance functions as well. In this article, we consider distance functions of the form $\varphi_{h,\psi}(\xvec,\yvec) = h\bigl\{d^{-1} \sum_{q=1}^d \psi(|x^{(q)} - y^{(q)}|)\bigr\}$, where $h:{\mathbb R}_+\rightarrow {\mathbb R}_+$ and $\psi:{\mathbb R}_+\rightarrow {\mathbb R}_+$ are continuous, monotonically increasing functions with $h(0)=\psi(0)=0$ such that $\varphi_{h,\psi}$ is a distance in ${\mathbb R}^d$. Clearly, this class of distance functions include all $\ell_p$ distances (upto a scalar constant) with $p\ge 1$. We define the general version of MADD as
\vspace{-0.05in}
\begin{equation}\label{eqn::MADD}
\rho_{h,\psi}(\xvec,\yvec) = \frac{1}{n-2} \sum_{\zvec \in {\cal X} \setminus \{\xvec,\yvec\}} \bigl|\varphi_{h,\psi}(\xvec,\zvec) - \varphi_{h,\psi}(\yvec,\zvec)\bigr|.
\vspace{-0.05in}
\end{equation}
Using $h(t) = \sqrt{t}$ and $\psi(t) = t^2$, we get $\rho_{h,\psi}(\xvec,\yvec) = d^{-1/2} \rho(\xvec,\yvec)$, and we call it $\rho_0$. MADD has some nice properties as a dissimilarity index. Lemma~\ref{lemma::semi-metric} shows that it is a semi-metric.
\vspace{-0.05in}
\begin{lemma}\label{lemma::semi-metric}
If ${\cal X}=\{\xvec_1,\ldots,\xvec_n\}$ contains $n\ge 3$ observations, $\rho_{h,\psi}$ is a semi-metric on ${\cal X}$.
\end{lemma}
\vspace{-0.1in}
MADD is not a metric since it is possible to get $\xvec \ne \yvec$ such that $\rho_{h,\psi}(\xvec,\yvec)=0$. But, when $\cal X$ consists of continuous random vectors, for $\xvec, \yvec \in {\cal X}$ and $\xvec \neq \yvec$, $\rho_{h,\psi}(\xvec,\yvec)>0$ holds with probability one. So, for all practical purposes, it behaves like a metric.

Since $\varphi_{h,\psi}$ satisfies the triangle inequality, one can show that $\rho_{h,\psi}(\xvec,\yvec) \le \varphi_{h, \psi}(\xvec,\yvec)$ for all $\xvec, \yvec \in {\mathbb R}^d$. Thus, closeness in terms of $\varphi_{h,\psi}$ (e.g., Euclidean distance) also indicates closeness in terms of MADD, but not the converse. In particular, for high dimensional data, unlike the Euclidean distance, MADD usually provides small dissimilarities among observations from the same population, and that helps us to develop better clustering algorithms. 

\vspace{-0.15in}
\subsection{High dimensional behavior of MADD}

\vspace{-0.05in}
To study the high dimensional behavior of $\rho_{h,\psi}$ and associated clustering algorithms in details, we assume that ${\cal X}$ consists of $n$ independent observations on the measurement vector $\Xvec=(X^{(1)},\ldots,X^{(d)})^{\top}$ coming from a mixture of $k_0$ populations, where $n_i = |{\cal X}_i| \ge 2$ for all $i=1,\ldots,k_0$. We also make the following assumption. 
\vspace{-0.1in}
\begin{enumerate}[$({A}1)$]
\item {\it For independent observations $\Xvec$ and $\Yvec$ from $i$-th and $j$-th populations $(1 \le i,j \le k_0)$, $d^{-1} \sum_{q=1}^d \bigl\{\psi(|X^{(q)} - Y^{(q)}|) - E \psi(|X^{(q)} - Y^{(q)}|) \bigr\} \overset{P}{\rightarrow} 0$ as $d \to \infty$.}
\end{enumerate}
\vspace{-0.1in}
This assumption regarding weak convergence of the sequence $\{\psi(|X^{(q)} - Y^{(q)}|): q \ge 1\}$ is pretty common in the HDLSS literature. A sufficient condition for $(A1)$ is $Var\bigl\{\sum_{q=1}^d \psi(|X^{(q)} - Y^{(q)}|)\bigr\} = {\bf o}(d^2)$. If the component variables are independent and identically distributed ($i.i.d.$) with $E\psi(|X^{(1)}-Y^{(1)}|)<\infty$, then $(A1)$ holds. For sequences of dependent and non-identically distributed random variables, we need some additional conditions. Several sufficient conditions have been used by many researchers. For instance, \cite{HMN05} assumed a $\rho$-mixing condition on the measurement variables and uniform boundedness of their fourth order moments to study the high dimensional behavior of some classifiers based on the Euclidean distance (i.e., when $\psi(t)=t^2$). \cite{JM09} used some slightly weaker conditions to establish the high dimensional consistency of their estimated principle component directions. Similar conditions were used by \cite{Ahn12} and \cite{BMG14} for high dimensional asymptotics. \cite{BMG15} derived some sufficient conditions for the weak convergence of $\{\psi(|X^{(q)} - Y^{(q)}|): q \ge 1\}$. Some sufficient conditions based on mixingales were derived by \cite{A88} and \cite{dJ95}.

Suppose that $\Xvec$ and $\Zvec$ are independent observations from the $i$-th and the $\ell$-th populations $(1 \le i,\ell \le k_0)$. Then, using $(A1)$ and the continuity of $h$, one gets $|\varphi_{h,\psi}(\Xvec,\Zvec) - \varphi^*_{h,\psi}(i,\ell)| \stackrel{P}{\rightarrow} 0$, where $\varphi^*_{h,\psi}(i,\ell) = h\bigl\{d^{-1} \sum_{q=1}^d E \psi(|X^{(q)} - Z^{(q)}|) \bigr\}$. So, if $\Xvec$ and $\Yvec$ are from $i$-th and $j$-th populations, we have $|\rho_{h,\psi}(\Xvec,\Yvec)-\rho^\ast_{h,\psi}(i,j)| \stackrel{P}{\rightarrow} 0$, where $\rho^\ast_{h,\psi}(i,j) = (n-2)^{-1} \big[(n_i-1)|\varphi_{h,\psi}^\ast(i,j) - \varphi_{h,\psi}^\ast(i,i)| + (n_j-1) |\varphi_{h,\psi}^\ast(i,j)-\varphi_{h,\psi}^\ast(j,j)| + \sum_{\ell \ne i,j} n_{\ell} |\varphi_{h,\psi}^\ast(i,\ell)-\varphi_{h,\psi}^\ast(j,\ell)|\big]$.
This is formally stated below.
\vspace{-0.075in}
\begin{lemma}\label{lemma:MADD_asymptotic}
Suppose that we have $n$ independent observations from $k_0$ populations satisfying $(A1)$. If $\Xvec$ and $\Yvec$ come from the $i$-th and the $j$-th $(1 \le i,j \le k_0)$ populations, respectively, and $h$ is continuous, then $|\rho_{h,\psi}(\Xvec,\Yvec) - \rho^\ast_{h,\psi}(i,j)| \overset{P}{\rightarrow} 0$ as $d \to \infty$.
\end{lemma}
\vspace{-0.1in}
Clearly, $\rho^\ast_{h,\psi}(i,j)=0$ if $i=j$ and $\rho^\ast_{h,\psi}(i,j)\ge 0$ for $i \neq j$. However, for good performance of clustering algorithms based on MADD, one would like to choose $h$ and $\psi$ such that $\rho^\ast_{h,\psi}(i,j) > 0$ for $i \neq j$. Lemma \ref{lemma:MADD_generalasymptotic} guides us to some suitable choices in this regard.
\vspace{-0.075in}
\begin{lemma}\label{lemma:MADD_generalasymptotic}
If $h$ and $\psi$ are strictly increasing, and $\psi^{\prime}(t)/t$ is a non-constant monotone function on $(0,\infty)$, then for any $i \neq j$, $\rho^\ast_{h,\psi}(i,j)=0$ if and only if the $i$-th and the $j$-th populations have the same one-dimensional marginals.
\end{lemma}
\vspace{-0.1in}
There are several choices of $\psi$ satisfying the properties mentioned in Lemma \ref{lemma:MADD_generalasymptotic} \citep[see, e.g.,][]{BF10,BMG15}. Some of them (e.g., $\psi(t)=t$, $\psi(t)=t/(1+t)$, $\psi(t)=1-e^{-t}$) lead to distance functions in ${\mathbb R}$. For such choices of $\psi$, it is enough to take $h(t)=t$ to make $\varphi_{h,\psi}$ a distance in ${\mathbb R}^d$. In these cases, we have $\rho^\ast_{h,\psi}(i,j)>0$ unless the two populations have identical marginal distributions. For the Euclidean distance (i.e., $\psi(t)=t^2$), $\psi$ does not satisfy the property mentioned in Lemma~\ref{lemma:MADD_generalasymptotic}. But Lemma~\ref{lemma:MADD_euclidasymptotic} shows that even in that case, $\rho^\ast_{h,\psi}(i,j)$ turns out to be
positive for a large class of examples.
\vspace{-0.075in}
\begin{lemma}\label{lemma:MADD_euclidasymptotic}
Let $\muvec_i$ and $\sigmat_i$ be the mean vector and the dispersion matrix of the $i$-th population $(i=1,\ldots,k_0)$, respectively. For $h(t)=\sqrt{t}$ and $\psi(t)=t^2$, 
$\rho^\ast_{h,\psi}(i,j)$ takes the value $0$ if and only if $\muvec_i=\muvec_j$ and $trace(\sigmat_i-\sigmat_j)=0$.
\end{lemma}
\vspace{-0.1in}
Lemmas~\ref{lemma:MADD_generalasymptotic} and \ref{lemma:MADD_euclidasymptotic} show that for suitable choices of $h$ and $\psi$, we usually have $\rho_{h,\psi}^{\ast}(i,j)>0$ for all values of $d$. So, it is reasonable to make the following assumption.
\vspace{-0.1in}
\begin{enumerate}[$({A}2)$]
\item {\it For every $1 \le i \ne j \le k_0$, $\liminf_{d \to \infty} \rho^\ast_{h,\psi}(i,j) > 0$.}
\end{enumerate}
\vspace{-0.1in}
Note that $(A2)$ holds in Examples A and B discussed in Section~\ref{sec::intro}. It says that the separation between two populations is not asymptotically  negligible. We will use this assumption for investigating the high dimensional behavior of MADD based algorithms.

\vspace{-0.15in}
\subsection{High dimensional behavior of MADD based clustering}

\vspace{-0.075in}
We know that AvgL begins with $n$ groups, each consisting of a single observation. At each step, it chooses two closest groups and merges them into a single one. To measure closeness, $\Delta(C_i,C_j)= {(|{C}_i||{C}_j|)}^{-1}\sum_{\zvec \in {C}_i, \wvec \in {C}_j}\|\zvec-\wvec\|$ is used as the distance between two groups ${C}_i$ and ${C}_j$. AvgL stops merging when the pairwise distance between any two groups is bigger than a certain threshold or a specified number of groups is attained. The final groups thus formed are considered as the estimated clusters. Note that in Example A, for $\Xvec_1,\Xvec_2 \stackrel{i.i.d.}{\sim} {\cal N}_d({\bf 0}_d,\sigma_1^2{\sigmat}_d)$ and $\Yvec_1,\Yvec_2 \stackrel{i.i.d.}{\sim} {\cal N}_d(\muvec_d,\sigma_2^2{\sigmat}_d)$, we have $\Pr(\|\Xvec_1-\Xvec_2\|<\|\Xvec_1-\Yvec_1\|<\|\Yvec_1-\Yvec_2\|) \rightarrow 1$ as $d \rightarrow \infty$ (see the discussion at the beginning of Section~\ref{sec::MADD}). So, for large values of $d$, after the first $49$ steps, all observations from the first population were merged into a single group, and in each subsequent step, one observation from the second population was added to it. As a result, when AvgL ended with two estimated clusters, one of them had a single observation from the second population, and the other had the rest of the observations. This led to a Rand index of $0.505$. A similar phenomenon occurred in Example B as well, where two of the three clusters estimated by AvgL had one observation each from the third population (i.e., ${\cal U}_d(2,2.5)$), while the third cluster contained the rest. As a result, the Rand index turned out to be $0.662$. The same phenomenon was observed even when single or complete linkage was used instead of AvgL.

We observed a diametrically opposite behavior for AvgL$_0$, the MADD version of AvgL based on $\rho_0$, where $\rho_0$ is used in place of $\|\cdot\|$ to define $\Delta(C_i,C_j)$. In Example A, as $d \rightarrow \infty$, both $\rho_0(\Xvec_1,\Xvec_2)$ and $\rho_0(\Yvec_1,\Yvec_2)$ converge to $0$, while $\rho_0(\Xvec_1,\Yvec_1)$ converges to a positive constant. So, any linkage method based on $\rho_0$ leads to perfect clustering (i.e., zero Rand index) as $d$ increases. Similar phenomenon occurs in Example B as well. This property of AvgL($h,\psi$), the MADD version of AvgL based on $\rho_{h, \psi}$, is asserted by the following theorem.

\vspace{-0.075in}
\begin{theorem}\label{thm::exact cluster hierarchical}
Suppose that we have independent observations from $k_0$ populations satisfying $(A1)$. If $h$ and $\psi$ satisfy $(A2)$, and AvgL$(h,\psi)$ is used to estimate $k_0$ clusters in the data, its Rand index converges to $0$ in probability as $d$ tends to infinity.
\end{theorem}
\vspace{-0.1in}

For a given $k$, kM algorithm aims at finding $k$ groups $C_1,\ldots,C_k$ with centers $\mvec_1,\ldots,\mvec_k$ such that $\Phi(C_1,\ldots,C_k) = \sum_{j=1}^{k} \sum_{i:\xvec_i \in C_j} \|\xvec_i - \mvec_j\|^2$
is minimized. In practice, it starts with an initial choice of $k$ groups, and then at each step an observation $\xvec$ is assigned to the group having the center closest to it. Group centers are updated accordingly. This iterative process is terminated when no groups are modified further. Using the convergence results for Euclidean distance, one can show that in Examples A and B, for large $d$, $\Phi$ is minimized when we have the same type of estimated clusters as obtained by AvgL. Since $\Phi(C_1,\ldots,C_k) = \sum_{r=1}^{k} {(2|C_r|)}^{-1}\sum_{\zvec, \wvec \in C_r} \|\zvec-\wvec\|^2$, for the MADD version of kM (denoted by kM$(h,\psi)$), we minimize $\Phi^\ast(C_1,\ldots,C_k) = \sum_{r=1}^{k} {(2|C_r|)}^{-1}\sum_{\zvec, \wvec \in C_r} \rho_{h,\psi}^2(\zvec,\wvec)$. Again we start with $k$ initial groups and use an iterative algorithm. At each step, distance of an observation $\xvec$ from a group $C_j$ is computed as $\rho_{h,\psi}^{(0)}(\xvec,C_j)={|C_j|}^{-1} \sum_{\zvec \in C_j}\rho_{h,\psi}^2(\xvec,\zvec)$, and it is assigned to the group $C_{\tilde{k}}$, where $\tilde{k}= \argmin_j \rho_{h,\psi}^{(0)}(\xvec,C_j)$. This is done for all observations, and the process is repeated until convergence. From Lemma~\ref{lemma:MADD_asymptotic}, it is clear that for $k_0$ estimated clusters $C_1,\ldots,C_{k_0}$, $\Phi^\ast(C_1,\ldots,C_{k_0})$ attains its minimum value if and only if we have perfect clustering. So, kM$(h,\psi)$ had excellent performance in Examples A and  B, specially for large $d$. This perfect clustering property of kM$(h,\psi)$ is asserted by the following theorem.

\vspace{-0.075in}
\begin{theorem}\label{thm::exact cluster kmeans}
Suppose that we have independent observations from $k_0$ populations satisfying $(A1)$. If $h$ and $\psi$ satisfy $(A2)$, and kM$(h,\psi)$ is used to estimate $k_0$ clusters in the data, its Rand index converges to $0$ in probability as $d$ tends to infinity.
\end{theorem}
\vspace{-0.1in}

Theorems \ref{thm::exact cluster hierarchical} and \ref{thm::exact cluster kmeans} show the perfect clustering  property of AvgL$(h,\psi)$ and kM$(h,\psi)$ when $\liminf_{d \to \infty} \rho^\ast_{h,\psi}(i,j) > 0$ for all $i \neq j$. This holds when $\liminf_{d \to \infty} d^{-1} \sum_{q=1}^d \big\{2E\psi(|X^{(q)}-Y^{(q)}|) - E\psi(|X_1^{(q)}-X_2^{(q)}|) - E\psi(|Y_1^{(q)}-Y_2^{(q)}|)\big\} > 0$, where $\Xvec_1,\Xvec_2$ are from $i$-th population and $\Yvec_1,\Yvec_2$ are from $j$-th population (see the proof of Lemma \ref{lemma:MADD_generalasymptotic}). So, in some sense, $(A2)$ assumes that the total signal increases at least at the order of $d$. As pointed out by one of the reviewers, this is quite restrictive in practice. We can relax this condition if we make a slightly stronger assumption on $h$. Let us assume that for any pair of independent observations $\Xvec$ and $\Zvec$, $Var\big\{\sum_{q=1}^d \psi(|X^{(q)}-Z^{(q)}|)\big\}={\bf O}(\vartheta^2(d))$. Then the perfect clustering property of AvgL$(h, \psi)$ and kM$(h, \psi)$ can be proved under the following assumption.
\vspace{-0.1in}
\begin{enumerate}[$({A}2^{\circ})$]
\item {\it For every $i \ne j$, $\rho_{h,\psi}^\ast(i,j)\; d/ \vartheta(d) \to \infty$ as $d \to \infty$.}
\end{enumerate}
\vspace{-0.1in}
Note that if the component variables are $i.i.d.$ with $E \psi^2(|X^{(1)}-Z^{(1)}|)<\infty$, then $\vartheta(d) \asymp d^{1/2}$ (i.e., $\vartheta(d)$ and $d^{1/2}$ are of the same asymptotic order). Under appropriate moment condition, we have the same asymptotic order of $\vartheta(d)$ for $m$-dependent sequence of random variables as well. Also, under weak mixing conditions on the component variables, we have $\vartheta(d) = {\bf o}(d)$ \citep[see, e.g.,][Chap.~2]{LinLu96}. In all such situations, $d/\vartheta(d) \to \infty$, and hence $(A2)$ implies $(A2^\circ)$. Theorem \ref{thm::MADD_separation_CLT} shows the perfect clustering property of AvgL$(h,\psi)$ and kM$(h,\psi)$ under this weaker assumption $(A2^\circ)$ when $h$ is Lipschitz continuous.

\vspace{-0.05in}
\begin{theorem}\label{thm::MADD_separation_CLT}
Suppose that we have independent observations from $k_0$ populations satisfying $(A2^\circ)$. Also assume that $h$ is Lipschitz continuous and $\psi^\prime(t)/t$ is a non-constant monotone function. Then, Rand indices of AvgL$(h,\psi)$ and kM$(h,\psi)$ converge to $0$ as $d$ tends to infinity.
\end{theorem}
\vspace{-0.125in}

If $\Xvec$ and $\Yvec$ are two independent observations from the $i$-th and the $j$-th populations, under Lipschitz continuity of $h$, we have $\rho_{h,\psi}(\Xvec,\Yvec)=\rho_{h,\psi}^{*}(i,j)+{\bf O}_p(\vartheta(d)/d)$ (see the proof of Theorem \ref{thm::MADD_separation_CLT}). While $\vartheta(d)/d$ can be interpreted as the order of stochastic variation (noise), $\rho_{h,\psi}^\ast(i,j)$ can be viewed as the signal. Theorem \ref{thm::MADD_separation_CLT} shows the perfect clustering property of AvgL$(h, \psi)$ and kM$(h, \psi)$ when this signal-to-noise ratio diverges. Similar results may hold even when $h$ is not Lipschitz continuous. For instance, in the case of $\rho_0$, where $h(t)=\sqrt{t}$ does not satisfy the Lipschitz condition, we have the following result.

\vspace{-0.05in}
\begin{theorem}
\label{thm::MADD_Euclid_CLT}
Suppose that we have independent observations from $k_0$ populations, where the $i$-th $(i=1,\ldots,k_0)$ population has mean $\muvec_i$ and dispersion matrix $\sigmat_i$ that satisfies $\liminf_{d \to \infty} tr(\sigmat_i)/\vartheta(d) > 0$. For every $i \neq j$, if $\|\muvec_i-\muvec_j\|^2/\vartheta(d) \to \infty$ and/or $|tr(\sigmat_i)-tr(\sigmat_j)|/\vartheta(d) \to \infty$, then Rand indices of AvgL$_0$ and kM$_0$ converge to $0$ as $d \to \infty$.
\end{theorem}
\vspace{-0.125in}

Therefore, if $\vartheta(d) \asymp d^{1/2}$ (i.e., in cases of weak dependence among component variables), the perfect clustering property of AvgL$_0$ and kM$_0$ holds when $d^{-1/2}\|\muvec_i-\muvec_j\|^2 \to \infty$ and/or $d^{-1/2} |tr(\sigmat_i)-tr(\sigmat_j)| \to \infty$ as $d \to \infty$. 

The spectral clustering algorithm of \cite{SM00} also failed to perform well in Examples A and B considered in Section \ref{sec::intro}. Note that spectral clustering methods deal with an edge-weighted graph with nodes $\{\xvec_1,\ldots,\xvec_n\}$ and a symmetric weight matrix $\mathcal{W}=((w_{ij}))_{d \times d}$, where $w_{ij}$ represents similarity between $\xvec_i$ and $\xvec_j$. The matrix $\mathcal{W}$ is usually computed from a similarity matrix ${\cal S}$, and different methods are available for it \citep[see][]{L07}. Often ${\cal S}=((s_{ij}))$ itself is used as $\mathcal{W}$, and one popular choice is the radial basis function $s_{ij} = \exp\{-\|\xvec_i-\xvec_j\|^2/{2\sigma^2}\}$, where $\sigma$ is a tuning parameter that controls the degree of similarity. These algorithms implicitly  assume that $s_{ij}$ will be large (respectively, small) if $\xvec_i$ and $\xvec_j$ belong to the same population (respectively, different populations). Since that was not the case in 
Examples A and B, Spect had poor performance. However, we do not have this problem if $s_{ij}$ is defined using $\rho_{h,\psi}^2(\xvec_i,\xvec_j)$ instead of $\|\xvec_i-\xvec_j\|^2$. So, Spect$(h,\psi)$, spectral clustering based on $\rho_{h,\psi}$, is expected to perform well, especially for large values of $d$. We observed the same for Spect$_0$ (spectral clustering based on $\rho_0$) in Examples A and B.

MDP clustering algorithm \citep{Ahn12} largely depends on the data piling property, which occurs only when the dimension exceeds the sample size. So, as expected, it performed poorly in both examples for smaller values of $d$. Surprisingly, it failed in Example B even when $d$ was large. A simple investigation explains this artifact. MDP clustering algorithm estimates the clusters by using binary splits at each step. For observations from two populations with mean vectors $\muvec_1,\muvec_2$ and dispersion matrices $\sigmat_1,\sigmat_2$ satisfying $d^{-1} \|\muvec_1-\muvec_2\|^2 \rightarrow \nu_{12}$, $d^{-1} trace(\sigmat_1) \rightarrow \sigma_1^2$  and $d^{-1} trace(\sigmat_2) \rightarrow \sigma_2^2$ as $d \to \infty$, this algorithm perfectly separates the observations in the HDLSS set up when
\vspace{-0.05in}
\begin{equation} \label{eqn::MDP}
\nu_{12} + \frac{\sigma_1^2}{n_1} + \frac{\sigma_2^2}{n_2} > \min \Big\{\frac{G+n_1}{Gn_1}\sigma_1^2, \frac{G+n_2}{Gn_2}\sigma_2^2\Big\},
\end{equation}
\vspace{-0.05in}
where $G$ is a pre-specified minimum number of observations in a cluster. Following \cite{Ahn12}, we used $G=5$ throughout this article. Recall that in Example B, all three populations had the same location, and condition \eqref{eqn::MDP} was violated for each pair of populations. 

\vspace{-0.2in}
\subsection{Comparison of clustering algorithms using simulated datasets}\label{sec::simulation-cluster}

\vspace{-0.1in}
We analyzed some simulated data sets for further evaluation of different clustering algorithms. In each example, we generated the data set by taking $50$ observations from each population, and different algorithms were used on these data sets assuming the number of clusters to be known. For these examples, we considered $d=100, 200$ and $500$, and each experiment was repeated $100$ times. Average Rand indices of different algorithms were computed over these $100$ trials, and they are reported in Tables~\ref{table::simulation rand} and \ref{table::Rand_7+8}. MDP clustering needs the number of eigen-vectors $T$ to be specified. We used $T=1,2,3$ as in \cite{Ahn12}, and reported the best results. For MADD, we used $h(t)=\sqrt{t}, \psi(t)=t^2$; $h(t)=t, \psi(t)=t$; and $h(t)=t, \psi(t)=1-e^{-t}$. The MADD indices for these three cases will be denoted by $\rho_0$, $\rho_1$ and $\rho_2$, respectively. These three choices led to similar results in Examples~1--6 (descriptions are given below). So, in Table~\ref{table::simulation rand}, results are reported for $\rho_0$ only.

{\bf Example-1}: Observations were generated from three Gaussian distributions with the same scatter matrix ${\sigmat}^{\circ}_d=((0.5^{|i-j|}))_{d \times d}$  but different mean vectors $\muvec_1$, $\muvec_2$ and $\muvec_3$, respectively. We took $\muvec_1={\bf 0}_d$, while $\muvec_2$ (respectively, $\muvec_3$) had the first $d/2$ elements equal to $0.75$ (respectively, $-0.75$) and the rest equal to $0$.

{\bf Example-2}: We used observations from four normal distributions, ${\cal N}_d(\alphavec,{\sigmat}_d^{\circ})$, ${\cal N}_d(\betavec,4{\sigmat}_d^{\circ})$, ${\cal N}_d(-\alphavec,{\sigmat}_d^{\circ})$ and ${\cal N}_d(-\betavec,4{\sigmat}_d^{\circ})$, which differed in their locations and scales. The mean vector $\alphavec=(\alpha_1,\ldots,\alpha_d)^\top$ had $\alpha_i=1$ and $\alpha_i=0.5$ for even and odd values of $i$, respectively. We took $\betavec=(\beta_1,\ldots,\beta_d)^\top$ with $\beta_i=(-1)^{i}\alpha_i$ for $i=1,\ldots,d$, and ${\sigmat}_d^{\circ}$ as in Example-1.

\begin{figure}[h!]
\centering
\includegraphics[height=2.20in,width=5.50in]{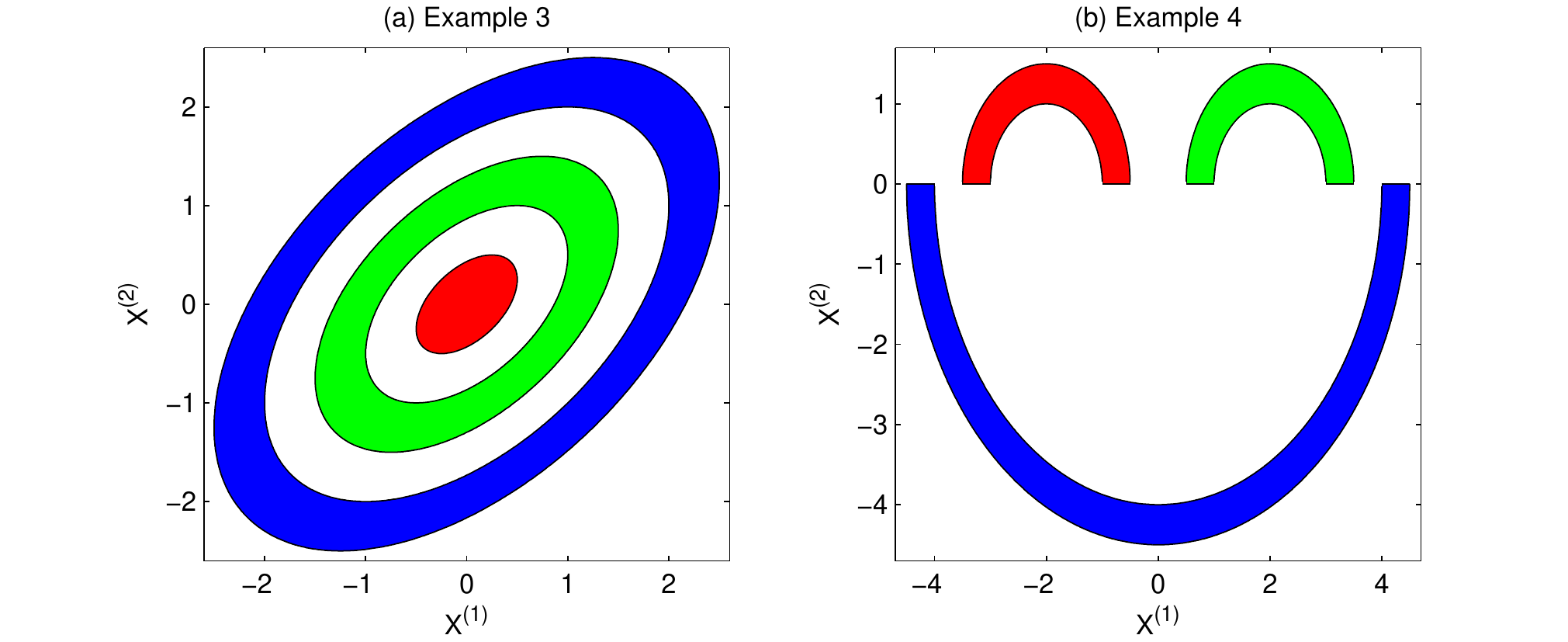}

\vspace{-0.2in}
\caption{Different populations in Examples 3 and 4 when $d=2$.}\label{fig:Example3&4}
\vspace{-0.15in}
\end{figure}

{\bf Example-3}: We considered three uniform distributions with disjoint supports $S_1$, $S_2$ and $S_3$, where $S_i=\{\xvec \in {\mathbb R}^{d}: ~i-1 \le \xvec^\prime{{\sigmat}_d^{\circ}}^{-1}\xvec \le i-1/2\}$ for $i=1,2,3$, and ${\bf \Sigma}_d^\circ$ is as in Example-1. Figure~\ref{fig:Example3&4}(a) shows the supports of the three distributions for $d=2$.

{\bf Example-4}: Define three sets $S_1^{\circ}=\{(x,y): y \ge 0, 1 \le \sqrt{(x-2)^2+y^2} \le 1.5\}$, $S_2^{\circ}=\{(x,y): y \ge 0, 1 \le \sqrt{(x+2)^2+y^2} \le 1.5\}$ and $S_3^{\circ}=\{(x,y): y \le 0, 4 \le \sqrt{x^2+y^2} \le 4.5\}$ (see Figure~\ref{fig:Example3&4}(b)). We generated $d/2$ independent observations from the uniform distribution on $S_i^{\circ}$ to get $d$ components of an observation from the $i$-th population ($i=1,2,3$).

{\bf Example-5}: Observations were generated from two auto-regressive processes $X^{(t)} = 0.75 + 0.25 X^{(t-1)} + \varepsilon_t$ and $X^{(t)} = 0.25 + 0.75 X^{(t-1)} + \varepsilon_t$ for $t=1,\ldots,d$. In both cases, we had $\varepsilon_t \sim {\cal N}(0,1)$ for every $t$. The distribution of $X^{(0)}$ was taken to be ${\cal N}(1,{16}/{15})$ and ${\cal N}(1,{16}/{7})$ in these two cases to make the processes stationary.

{\bf Example-6}: Let ${\cal S}_d$ be the $d$-dimensional unit sphere with center at the origin, and ${\cal C}_d$ be the largest hypercube inscribed in it. We considered two uniform distributions, one on ${\cal S}_d$ and the other on ${\cal C}_d$. Note that if $\Xvec$ comes from the first population, then $\Pr(\Xvec \in {\cal C}_{d}) \rightarrow 0$ as $d \rightarrow \infty$. So, the two populations become completely separated in high dimensions.

\begin{table}[h!]
\begin{center}
\spacingset{1}
\setlength{\tabcolsep}{5.0pt}
\caption{Average Rand indices of different clustering algorithms in Examples 1--6}
\label{table::simulation rand}
\vspace{-0.1in}
\small
\begin{tabular}{|c|c|c|c|c|c|c|c|c|} \hline
 & $d$ & AvgL & AvgL$_0$ & kM & kM$_0$ & MDP & Spect & Spect$_0$ \\ \hline
     & $100$ & $0.2906$ & $0.0865$ & ${\bf 0.0185}$ & $0.0367$ & $0.5801$ & $0.2512$ & $0.1851$ \\
Ex-1 & $200$ & $0.2168$ & $0.0104$ & $0.0201$ & $0.0095$ & ${\bf 0.0000}$ & $0.2419$ & $0.1953$ \\
     & $500$ & $0.0429$ & ${\bf 0.0000}$ & $0.0074$ & ${\bf 0.0000}$ & ${\bf 0.0000}$ & $0.2330$ & $0.1919$ \\ \hline
     & $100$ & $0.7335$ & $0.0502$ & $0.4206$ & ${\bf 0.0067}$ & $0.6204$ & $0.2609$ & $0.0415$ \\
Ex-2 & $200$ & $0.7361$ & $0.0115$ & $0.6206$ & ${\bf 0.0001}$ & $0.0714$ & $0.2462$ & $0.0440$ \\
     & $500$ & $0.7378$ & ${\bf 0.0000}$ & $0.6982$ & ${\bf 0.0000}$ & $0.0018$ & $0.2187$ & $0.0434$ \\ \hline
     & $100$ & $0.6616$ & ${\bf 0.0000}$ & $0.6608$ & ${\bf 0.0000}$ & $0.5885$ & $0.4492$ & $0.2348$ \\
Ex-3 & $200$ & $0.6619$ & ${\bf 0.0000}$ & $0.6617$ & ${\bf 0.0000}$ & $0.5826$ & $0.4513$ & $0.2298$ \\
     & $500$ & $0.6619$ & ${\bf 0.0000}$ & $0.6619$ & ${\bf 0.0000}$ & $0.5761$ & $0.4515$ & $0.2286$ \\ \hline
     & $100$ & $0.2346$ & ${\bf 0.0000}$ & $0.2762$ & ${\bf 0.0000}$ & $0.5841$ & $0.0745$ & $0.0417$ \\
Ex-4 & $200$ & $0.2330$ & ${\bf 0.0000}$ & $0.2769$ & ${\bf 0.0000}$ & ${\bf 0.0000}$ & $0.0714$ & $0.0361$ \\
     & $500$ & $0.2327$ & ${\bf 0.0000}$ & $0.2748$ & ${\bf 0.0000}$ & ${\bf 0.0000}$ & $0.0752$ & $0.0478$ \\ \hline
     & $100$ & $0.5047$ & $0.3516$ & $0.5027$ & ${\bf 0.2271}$ & $0.4895$ & $0.4801$ & $0.2584$ \\
Ex-5 & $200$ & $0.5048$ & ${\bf 0.0762}$ & $0.5040$ & $0.0784$ & $0.4863$ & $0.4813$ & $0.1231$ \\
     & $500$ & $0.5048$ & ${\bf 0.0028}$ & $0.5048$ & $0.0060$ & $0.4795$ & $0.4726$ & $0.0119$ \\ \hline
     & $100$ & $0.5047$ & ${\bf 0.0000}$ & $0.5042$ & ${\bf 0.0000}$ & $0.4857$ & $0.5000$ & ${\bf 0.0000}$ \\
Ex-6 & $200$ & $0.5048$ & ${\bf 0.0000}$ & $0.5048$ & ${\bf 0.0000}$ & $0.4836$ & $0.4992$ & ${\bf 0.0000}$ \\
     & $500$ & $0.5048$ & ${\bf 0.0000}$ & $0.5048$ & ${\bf 0.0000}$ & $0.4803$ & $0.4992$ & ${\bf 0.0000}$ \\ \hline
\end{tabular}

\vspace{0.05in}
{\footnotesize Bold figures indicate the best result in each example.}
\end{center}
\vspace{-0.25in}
\end{table}

Table~\ref{table::simulation rand} clearly shows that both AvgL$_0$ and kM$_0$ performed much better than AvgL and kM. In all six examples, they led to perfect clustering, especially for higher values of $d$. MDP clustering algorithm performed poorly for $d=100$. For $d=200$ and $500$, it performed well in Examples 1, 2 and 4, but in the other three examples, where the population distributions did not differ in their means, it had miserable performance. Spect$_0$ also performed better than Spect. In Example-6, when all other clustering algorithms had Rand indices close to $0.5$, those based on $\rho_0$ led to perfect clustering for all values of $d$ considered here. Among them, overall performance of AvgL$_0$ and kM$_0$ was much better than Spect$_0$.

Next, we considered two examples, where clustering based on $\rho_0$, $\rho_1$ and $\rho_2$ led to widely varying results (see Table~\ref{table::Rand_7+8}). Descriptions of these two data sets are given below.

{\bf Example-7}: Observations were generated from four normal distributions having the same mean ${\bf 0}_d$ and diagonal dispersion matrices. For the first (respectively, second) population, the first $d/2$ diagonal elements were $1$ (respectively, $9$) and the rest were $9$ (respectively,  $1$). The scatter matrix of the third (respectively, fourth) population had $1$ and $9$ (respectively, $9$ and $1$) at even and odd places along the diagonal, respectively.

{\bf Example-8}: We considered two populations where all the measurement variables were $i.i.d$. For the first population, they were distributed as ${\cal N}(0,3)$, while they had standard $t_3$
($t$ with 3 d.f) distribution for the second population. So, the two populations had the same mean vector and dispersion matrix, but they differed in their shapes.

\begin{table}[h]
\centering
\renewcommand{\tabcolsep}{0.075cm}
\caption{Average Rand indices of different clustering algorithms in Examples 7 and 8}
\label{table::Rand_7+8}
\vspace{-0.1in}
\footnotesize
\begin{tabular}{|c|c|c|c|c|c|c|c|c|c|c|c|c|c|c|} \hline
 & $d$ & AvgL & AvgL$_0$ & AvgL$_1$& AvgL$_2$& kM & kM$_0$ & kM$_1$& kM$_2$& MDP & {Spect} & {Spect}$_0$
 & {Spect}$_1$& {Spect}$_2$\\ \hline
 \parbox[t]{3mm}{\multirow{3}{*}{\rotatebox[origin=c]{90}{Ex-7}}}    & $100$ & 0.7366 & 0.4831 & 0.4914 & 0.0044 & 0.4432 & 0.4102 & 0.2721 & {\bf 0.0001} & 0.5868 & 0.7034 & 0.3765 & 0.1732 & 0.0907 \\
     & 200 & 0.7364 & 0.4873 & 0.3168 & 0.0001 & 0.4593 & 0.4082 & 0.0935 & {\bf 0.0000} & 0.6252 & 0.6990 & 0.3767 & 0.1310 & 0.0646 \\
     & 500 & 0.7370 & 0.4776 & 0.0471 & {\bf 0.0000} & 0.4522 & 0.4048 & 0.0192 & {\bf 0.0000} & 0.6173 & 0.6975 & 0.3756 & 0.0903 & 0.0540 \\ \hline
 \parbox[t]{3mm}{\multirow{3}{*}{\rotatebox[origin=c]{90}{Ex-8}}}
    & $100$ & 0.5048 & 0.5020 & 0.3883 & 0.1309 & 0.5048 & 0.4955 & 0.3132 & \bf 0.0845 & 0.5021 & 0.5049 & 0.4894 & 0.3127 & 0.0956 \\
    & 200 & 0.5048 & 0.5021 & 0.2837 & 0.0251 & 0.5048 & 0.4930 & 0.2087 & \bf 0.0157 & 0.5027 & 0.5048 & 0.4801 & 0.2138 & 0.0188 \\
    & 500 & 0.5049 & 0.5003 & 0.1109 & 0.0002 & 0.5049 & 0.4888 & 0.0889 & \bf 0.0000 & 0.5029 & 0.5047 & 0.4818 & 0.0878 & \bf 0.0000 \\ \hline
\end{tabular}

\vspace{0.05in}
{\footnotesize Bold figures indicate the best result in each example.}
\vspace{-0.05in}
\end{table}

Table~\ref{table::Rand_7+8} shows that AvgL, kM, MDP and Spect, all had miserable performance in these two examples. Even MADD clustering algorithms failed when $\rho_0$ was used, but those based on $\rho_1$ (denoted by AvgL$_1$, kM$_1$ and Spect$_1$) and $\rho_2$ (denoted by AvgL$_2$, kM$_2$ and Spect$_2$) had improved performance. In these examples, we have $\rho_{h,\psi}^{*}(i,j)=0$ for all $i \ne j$ when $\rho_0$ is used, but they are positive for $\rho_1$ and $\rho_2$. That was the reason for their improved performance. Among these two choices, $\rho_2$, which is based on a bounded $\psi$ function, yielded better results. We also observed similar phenomenon when the $t$ distribution in Example-8 was replaced by the standard Cauchy distribution. In that case, clustering algorithms based on $\rho_2$ had Rand indices close to $0$ for all choices of $d$, but those for all other methods were close to $0.5$. This shows the robustness of MADD clustering algorithms based on bounded $\psi$ functions against heavy tailed distributions.

\vspace{-0.175in}
\section{Estimation of the number of clusters}\label{sec::no-of-clusters}

\vspace{-0.075in}
So far we have assumed $k_0$ to be known for our analysis. But in practice, one needs to estimate $k_0$. Several estimation methods have been proposed for it \sloppy\citep[see][]{CH74,H75,KL85, KR90, TWH01, SJ03,W10}. Brief descriptions of some of these  methods, that we use in this article, are given below.
{These estimation methods can be used with any base clustering algorithm}. Throughout this article, we use average linkage or $k$-means algorithm (either based on Euclidean distance or based on MADD) for base clustering. 

\textsc{KL} statistic \citep{KL85}: For a given $k$, if $C_1,\ldots,C_k$ are the $k$ clusters estimated by the base clustering algorithm, then the \textsc{KL} statistic is defined as $\textsc{KL}(k) = \big|{\textsc{Diff}(k)}/{\textsc{Diff}(k+1)}\big|$, where $\textsc{Diff}(k) = (k-1)^{2/d}W_{k-1}-k^{2/d}W_k$, and $W_{k}= \sum_{j=1}^{k} {(2|C_j|)}^{-1} \sum_{\zvec,\wvec \in C_j} \|\zvec-\wvec\|^2$ is the within group sum of squares. $\textsc{KL}(k)$ is computed for a range of values $\{2,\ldots,K\}$ of $k$ and ${\hat k}_{KL}=\argmax_{2 \le k \le K} \textsc{KL}(k)$ is used to estimate $k_0$.

\textsc{Gap} statistic \citep{TWH01}: For any fixed $k$, the \textsc{Gap} statistic is defined as $\textsc{Gap}(k) ={B}^{-1} \sum_{b=1}^B \log(W_{k}^{(b)})-\log(W_k)$, where $W_{k}$ is as defined above, and $W_{k}^{(b)}$ is the within group sum of squares computed using the $b$-th bootstrap sample ($b=1,\ldots,B$) generated from a reference distribution. The number of clusters $k_0$ is estimated by ${\hat k}_{G}=\min\{k:~\textsc{Gap}(k) \ge \textsc{Gap}(k+1)-s_{k+1}\}$, where $s_k = \sqrt{(1+{B}^{-1})}sd_k$, and $sd_k$ is the standard deviation of $\log(W_{k}^{(b)})$. Unlike the {\sc KL} statistic, ${\textsc{Gap}}(k)$ can be defined for $k=1$ as well.

\textsc{Jump} statistic \citep{SJ03}: For $k\ge 1$, the \textsc{Jump} statistic is defined as $\textsc{Jump}(k)=\hat{d}_k^{-t} - \hat{d}_{k-1}^{-t}$, where $\hat{d}_0^{-t}=0$,  $\hat{d}_k = {d}^{-1} \sum_{i=1}^n \min_{r=1,\ldots,k} (\xvec_i-\mvec_r)^\top \Gammavec^{-1}(\xvec_i-\mvec_r)$ for $k\ge 1$, and $\mvec_r$ is the center of the $r$-th cluster. The number of clusters is estimated by  ${\hat k}_{J}=\argmax_{1 \le k \le K} \textsc{Jump}(k)$. The authors suggested to use $\Gammavec = {\bf I}_d$ (the $d\times d$ identity matrix) and $t={d/2}$. Note that for $k$-means clustering with $\Gammavec={\bf I}_d$, we get  ${\hat d}_k={d}^{-1}\sum_{i=1}^n \min_{r=1,\ldots,k} \|\xvec_i-\mvec_r\|^2$ = ${d}^{-1} \sum_{j=1}^k \sum_{\zvec \in C_j} \|\zvec-\mvec_j\|^2$ = ${d}^{-1}\sum_{j=1}^k {(2|C_j|)}^{-1} \sum_{\zvec,\wvec \in C_j} \|\zvec-\wvec\|^2 = W_k/d$.

Cross-validated Rand index \citep{W10}: The whole sample ${\cal X}$ is randomly divided into three parts ${\cal X}^{(1)},{\cal X}^{(2)}$ and ${\cal X}^{(3)}$ of sizes $m,m$ and $n-2m$, respectively. For any given $k$, the first two parts are used to develop two clustering algorithms $\delta_1=\delta_{{\cal X}^{(1)},k}$ and $\delta_2=\delta_{{\cal X}^{(2)},k}$, which are then used on ${\cal X}^{(3)}$ to estimate clustering instability given by $\textsc{Ins}(k) ={\binom{n-2m}{2}}^{-1} \sum_{\xvec \ne \yvec \in {\cal X}^{(3)}} \mathbb{I}\Bigl[\mathbb{I}\{\delta_1(\xvec)=\delta_1(\yvec)\}+\mathbb{I}\{\delta_2(\xvec)=\delta_2(\yvec)\}=1\Bigr]$. This process is repeated $B$ times, and the results are aggregated. The author proposed two methods for aggregation. In one method (call it $\textsc{CV}_a$), the average instability over $B$ repetitions is computed, and $k_0$ is estimated by minimizing this average instability with respect to $k$. In the other method (call it $\textsc{CV}_v$), for each repetition, the number of clusters is estimated by minimizing $\textsc{Ins}(k)$ over $k$, and finally the modal value of the minimizers is used as the estimator of $k_0$.

We use another method based on the {\sc Dunn} index \citep{D73}. For fixed $k$, let $C_1,\ldots,C_k$ be the clusters estimated by a base clustering algorithm. Define $\Delta_0(C_i) = \{|C_i|(|C_i|-1)\}^{-1}$ $\sum_{\zvec,\wvec \in C_i} \|\zvec-\wvec\|$ and  $\Delta(C_i,C_j) = {(|C_i||C_j|)}^{-1} \sum_{\zvec \in C_i, \wvec \in C_j} \|\zvec-\wvec\|$ (instead of average, other suitable measures can also be used). The {\sc Dunn} index is given by $\textsc{D}(k) = B_k^{\circ}/W_k^{\circ}$, where $W_k^{\circ}=\max_{1 \le i \le k} \Delta_0(C_i)$ and $B_k^{\circ}=\min_{1 \le i < j \le k} \Delta(C_i,C_j)$, respectively. \cite{D73} used this index for cluster validation, but here we use it to estimate $k_0$ by ${\hat k}_D= \argmax_{2 \le k \le K} D(k)$.

When AvgL or kM is used for base clustering, we use usual versions of these statistics. But, when AvgL($h, \psi$) or kM($h, \psi$) is used for base clustering, we use $\rho_{h,\psi}$ in place of $\|\cdot\|$ to define $W_k$ for \textsc{KL}, \textsc{Gap} and \textsc{Jump} statistics, and to define $\Delta_0, \Delta$  for the {\sc Dunn} index.

\begin{figure}[h!]
\spacingset{1}
\centering
\begin{subfigure}[b]{\linewidth}
\centering
\subcaption{Example 1 $(k_0=3)$.}
\vspace{-0.1in}
\includegraphics[width=6.0in,height=2.20in]{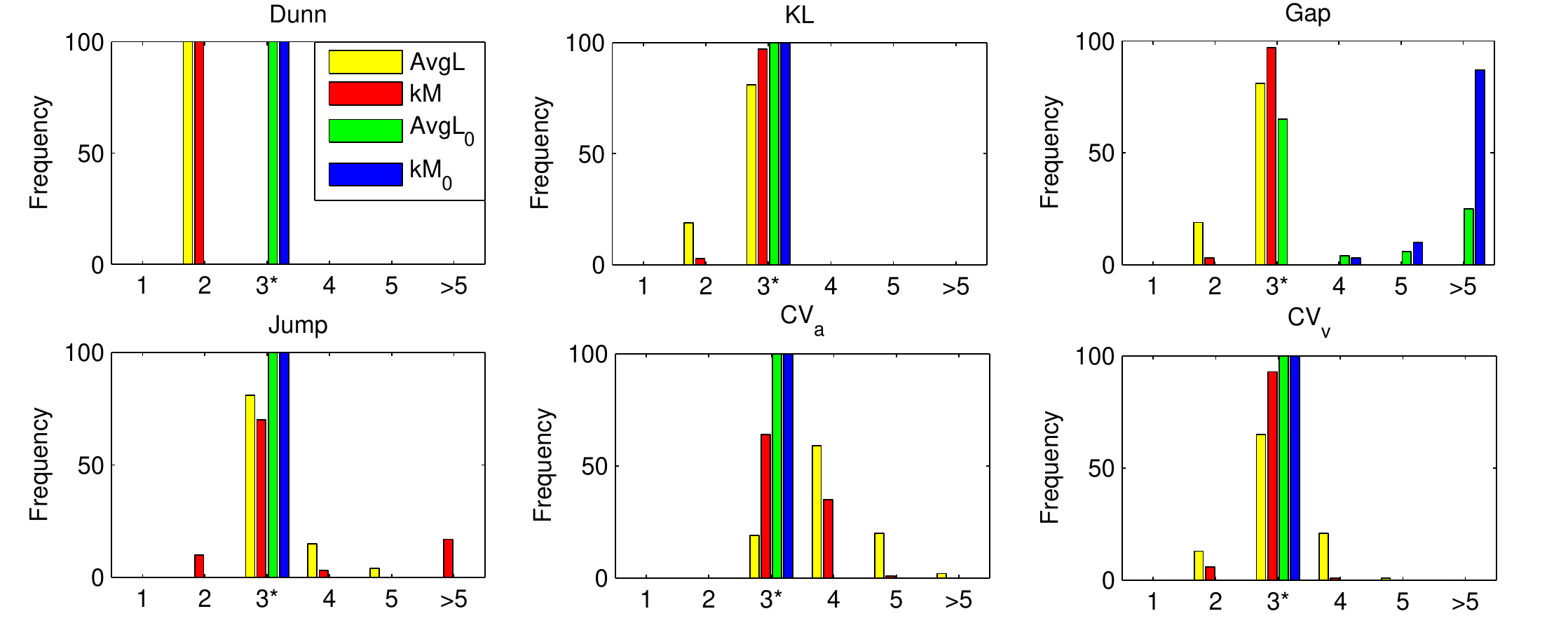}
\end{subfigure}
\vspace{-0.05in}
\begin{subfigure}[b]{\linewidth}
\centering
\subcaption{Example 2 $(k_0=4)$.}
\vspace{-0.1in}
\includegraphics[width=6.0in,height=2.20in]{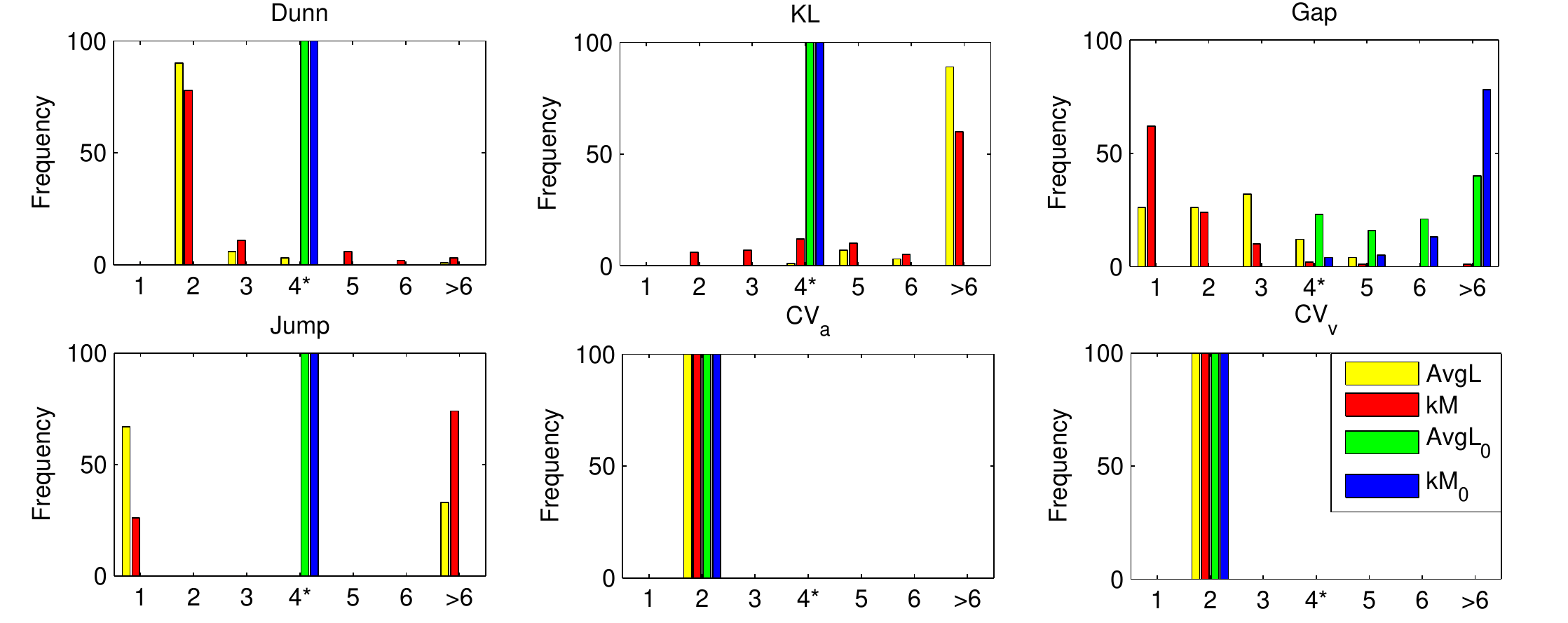}
\end{subfigure}

\caption{Barplots for $k_0$ estimated by different methods in Examples 1 and 2.}
\label{fig::bar_Ex3+4}
\vspace{-0.125in}
\end{figure}

To investigate the performance of these estimation methods, we considered the examples used in Section~\ref{sec::simulation-cluster} with $d=500$. For CV$_a$ and CV$_v$, we used $B=100$, and $m$ was taken to be the largest multiple of $5$ not exceeding ${n/3}$. For the \textsc{Gap} statistic, we used $B=100$, and bootstrap samples were generated from the uniform distribution on the range of the measurement variables. When $d$ was large (in the order of hundreds or more), the use of $t={d/2}$ led to poor performance by the \textsc{Jump} statistic. So, we tried several values of $t$, and based on our empirical experience, we selected $t=1$ when MADD was used. However, we were unable to find any such $t$ when the Euclidean norm was used. In such cases, we performed our experiments with several choices of $t$ and here we report the best results. Throughout this article, we consider values of $k$ in the range $\{1,\ldots,12\}$. However, only \textsc{Gap} and \textsc{Jump} statistics are defined for $k=1$.

Figure~\ref{fig::bar_Ex3+4} shows barplots for the number of clusters estimated by different methods in Examples 1 and 2.  From this figure, it is clear that barring the \textsc{Gap} statistic, all other methods worked better when $\rho_0$ was used. The results based on $\rho_0, \rho_1$ and $\rho_2$ were almost similar. We observed this phenomenon in Examples 3--6 as well. So, for reporting the performance of these methods in Examples 1--6, we considered the results based on $\rho_0$ only (see Table~\ref{table::simulation k}). For the \textsc{Gap} statistic, there was no clear winner. For this method, we used both the Euclidean distance and $\rho_0$ in all examples, and in each case, the best result Table~\ref{table::simulation k} clearly shows that except for Examples 1 and 6, \textsc{Gap} statistic performed poorly throughout. $\textsc{CV}_a$ and $\textsc{CV}_v$ also underestimated $k_0$ in Examples 2 and 3. But KL statistic, \textsc{Jump} statistic and {\sc Dunn} index correctly estimated $k_0$ on all occasions.

\begin{table}[h!]
\spacingset{1}
\setlength{\tabcolsep}{0.4mm}
\begin{center}
\caption{Frequency distribution for the estimated number of clusters in Examples 1--6}
\label{table::simulation k}
\vspace{-0.05in}
\footnotesize
\begin{tabular}{|c|c|ccccccccccccHHH|ccccccccccccHHH|} \hline
 & & \multicolumn{15}{c|}{AvgL} & \multicolumn{15}{c|}{kM} \\ \cline{2-32}
& $k$ & ~1~ & ~2~ & ~3~ & ~4~ & ~5~ & ~6~ & ~7~ & ~8~ & ~9~ & 10 & 11 & 12 & 13 & 14 & 15 & ~1~ & ~2~ & ~3~ & ~4~ & ~5~ & ~6~ & ~7~ & ~8~ & ~9~ & 10 & 11 & 12 & 13 & 14 & 15 \\ \hline

\parbox[t]{3mm}{\multirow{7}{*}{\rotatebox[origin=c]{90}{Ex-1}}}
& {\sc Dunn}$^\ast$ & $0$ & $0$ & ${\bf 100}$ & $0$ & $0$ & $0$ & $0$ & $0$ & $0$ & $0$ & $0$ & $0$ & $0$ & $0$ & $0$ & $0$ & $0$ & ${\bf 100}$ & $0$ & $0$ & $0$ & $0$ & $0$ & $0$ & $0$ & $0$ & $0$ & $0$ & $0$ & $0$ \\
& \textsc{KL}$^\ast$ & $0$ & $0$ & ${\bf 100}$ & $0$ & $0$ & $0$ & $0$ & $0$ & $0$ & $0$ & $0$ & $0$ & $0$ & $0$ & $0$ & $0$ & $0$ & ${\bf 100}$ & $0$ & $0$ & $0$ & $0$ & $0$ & $0$ & $0$ & $0$ & $0$ & $0$ & $0$ & $0$ \\
& \textsc{Gap}$^?$ & $0$ & $19$ & ${\bf 81}$ & $0$ & $0$ & $0$ & $0$ & $0$ & $0$ & $0$ & $0$ & $0$ & $0$ & $0$ & $0$ & $0$ & $3$ & ${\bf 97}$ & $0$ & $0$ & $0$ & $0$ & $0$ & $0$ & $0$ & $0$ & $0$ & $0$ & $0$ & $0$ \\
& \textsc{Jump}$^\ast$ & $0$ & $0$ & ${\bf 100}$ & $0$ & $0$ & $0$ & $0$ & $0$ & $0$ & $0$ & $0$ & $0$ & $0$ & $0$ & $0$ & $0$ & $0$ & ${\bf 100}$ & $0$ & $0$ & $0$ & $0$ & $0$ & $0$ & $0$ & $0$ & $0$ & $0$ & $0$ & $0$ \\
& $\textsc{CV}_a^\ast$ & $0$ & $0$ & ${\bf 100}$ & $0$ & $0$ & $0$ & $0$ & $0$ & $0$ & $0$ & $0$ & $0$ & $0$ & $0$ & $0$ & $0$ & $0$ & ${\bf 100}$ & $0$ & $0$ & $0$ & $0$ & $0$ & $0$ & $0$ & $0$ & $0$ & $0$ & $0$ & $0$ \\
& $\textsc{CV}_v^\ast$ & $0$ & $0$ & ${\bf 100}$ & $0$ & $0$ & $0$ & $0$ & $0$ & $0$ & $0$ & $0$ & $0$ & $0$ & $0$ & $0$ & $0$ & $0$ & ${\bf 100}$ & $0$ & $0$ & $0$ & $0$ & $0$ & $0$ & $0$ & $0$ & $0$ & $0$ & $0$ & $0$ \\ \hline
\parbox[t]{3mm}{\multirow{7}{*}{\rotatebox[origin=c]{90}{Ex-2}}}
& {\sc Dunn}$^\ast$ & $0$ & $0$ & $0$ & ${\bf 100}$ & $0$ & $0$ & $0$ & $0$ & $0$ & $0$ & $0$ & $0$ & $0$ & $0$ & $0$ & $0$ & $0$ & $0$ & ${\bf 100}$ & $0$ & $0$ & $0$ & $0$ & $0$ & $0$ & $0$ & $0$ & $0$ & $0$ & $0$ \\
& \textsc{KL}$^\ast$ & $0$ & $0$ & $0$ & ${\bf 100}$ & $0$ & $0$ & $0$ & $0$ & $0$ & $0$ & $0$ & $0$ & $0$ & $0$ & $0$ & $0$ & $0$ & $0$ & ${\bf 100}$ & $0$ & $0$ & $0$ & $0$ & $0$ & $0$ & $0$ & $0$ & $0$ & $0$ & $0$ \\
& \textsc{Gap}$^?$ & $0$ & $0$ & $0$ & ${\bf 23}$ & $16$ & $21$ & $15$ & $12$ & $5$ & $1$ & $5$ & $2$ & $0$ & $0$ & $0$ & $0$ & $0$ & $0$ & ${\bf 4}$ & $5$ & $13$ & $20$ & $14$ & $18$ & $10$ & $10$ & $6$ & $0$ & $0$ & $0$ \\
& \textsc{Jump}$^\ast$ & $0$ & $0$ & $0$ & ${\bf 100}$ & $0$ & $0$ & $0$ & $0$ & $0$ & $0$ & $0$ & $0$ & $0$ & $0$ & $0$ & $0$ & $0$ & $0$ & ${\bf 100}$ & $0$ & $0$ & $0$ & $0$ & $0$ & $0$ & $0$ & $0$ & $0$ & $0$ & $0$ \\
& $\textsc{CV}_a^\ast$ & $0$ & $100$ & $0$ & ${\bf 0}$ & $0$ & $0$ & $0$ & $0$ & $0$ & $0$ & $0$ & $0$ & $0$ & $0$ & $0$ & $0$ & $100$ & $0$ & ${\bf 0}$ & $0$ & $0$ & $0$ & $0$ & $0$ & $0$ & $0$ & $0$ & $0$ & $0$ & $0$ \\
& $\textsc{CV}_v^\ast$ & $0$ & $100$ & $0$ & ${\bf 0}$ & $0$ & $0$ & $0$ & $0$ & $0$ & $0$ & $0$ & $0$ & $0$ & $0$ & $0$ & $0$ & $100$ & $0$ & ${\bf 0}$ & $0$ & $0$ & $0$ & $0$ & $0$ & $0$ & $0$ & $0$ & $0$ & $0$ & $0$ \\ \hline
\parbox[t]{3mm}{\multirow{7}{*}{\rotatebox[origin=c]{90}{Ex-3}}}
& {\sc Dunn}$^\ast$ & $0$ & $0$ & ${\bf 100}$ & $0$ & $0$ & $0$ & $0$ & $0$ & $0$ & $0$ & $0$ & $0$ & $0$ & $0$ & $0$ & $0$ & $0$ & ${\bf 100}$ & $0$ & $0$ & $0$ & $0$ & $0$ & $0$ & $0$ & $0$ & $0$ & $0$ & $0$ & $0$ \\
& \textsc{KL}$^\ast$ & $0$ & $0$ & ${\bf 100}$ & $0$ & $0$ & $0$ & $0$ & $0$ & $0$ & $0$ & $0$ & $0$ & $0$ & $0$ & $0$ & $0$ & $0$ & ${\bf 100}$ & $0$ & $0$ & $0$ & $0$ & $0$ & $0$ & $0$ & $0$ & $0$ & $0$ & $0$ & $0$ \\
& \textsc{Gap}$^?$ & $0$ & $0$ & ${\bf 12}$ & $13$ & $41$ & $18$ & $13$ & $3$ & $0$ & $0$ & $0$ & $0$ & $0$ & $0$ & $0$ & $0$ & $0$ & ${\bf 0}$ & $2$ & $20$ & $35$ & $23$ & $18$ & $1$ & $1$ & $0$ & $0$ & $0$ & $0$ & $0$ \\
& \textsc{Jump}$^\ast$ & $0$ & $0$ & ${\bf 100}$ & $0$ & $0$ & $0$ & $0$ & $0$ & $0$ & $0$ & $0$ & $0$ & $0$ & $0$ & $0$ & $0$ & $0$ & ${\bf 100}$ & $0$ & $0$ & $0$ & $0$ & $0$ & $0$ & $0$ & $0$ & $0$ & $0$ & $0$ & $0$ \\
& $\textsc{CV}_a^\ast$ & $0$ & $100$ & ${\bf 0}$ & $0$ & $0$ & $0$ & $0$ & $0$ & $0$ & $0$ & $0$ & $0$ & $0$ & $0$ & $0$ & $0$ & $100$ & ${\bf 0}$ & $0$ & $0$ & $0$ & $0$ & $0$ & $0$ & $0$ & $0$ & $0$ & $0$ & $0$ & $0$ \\
& $\textsc{CV}_v^\ast$ & $0$ & $100$ & ${\bf 0}$ & $0$ & $0$ & $0$ & $0$ & $0$ & $0$ & $0$ & $0$ & $0$ & $0$ & $0$ & $0$ & $0$ & $100$ & ${\bf 0}$ & $0$ & $0$ & $0$ & $0$ & $0$ & $0$ & $0$ & $0$ & $0$ & $0$ & $0$ & $0$ \\ \hline
\parbox[t]{3mm}{\multirow{7}{*}{\rotatebox[origin=c]{90}{Ex-4}}}
& {\sc Dunn}$^\ast$ & $0$ & $0$ & ${\bf 100}$ & $0$ & $0$ & $0$ & $0$ & $0$ & $0$ & $0$ & $0$ & $0$ & $0$ & $0$ & $0$ & $0$ & $0$ & ${\bf 100}$ & $0$ & $0$ & $0$ & $0$ & $0$ & $0$ & $0$ & $0$ & $0$ & $0$ & $0$ & $0$ \\
& \textsc{KL}$^\ast$ & $0$ & $0$ & ${\bf 100}$ & $0$ & $0$ & $0$ & $0$ & $0$ & $0$ & $0$ & $0$ & $0$ & $0$ & $0$ & $0$ & $0$ & $0$ & ${\bf 100}$ & $0$ & $0$ & $0$ & $0$ & $0$ & $0$ & $0$ & $0$ & $0$ & $0$ & $0$ & $0$ \\
& \textsc{Gap}$^?$ & $0$ & $20$ & ${\bf 19}$ & $10$ & $15$ & $9$ & $9$ & $4$ & $9$ & $1$ & $2$ & $2$ & $0$ & $0$ & $0$ & $0$ & $3$ & ${\bf 27}$ & $23$ & $24$ & $9$ & $4$ & $6$ & $2$ & $0$ & $1$ & $1$ & $0$ & $0$ & $0$ \\
& \textsc{Jump}$^\ast$ & $0$ & $0$ & ${\bf 100}$ & $0$ & $0$ & $0$ & $0$ & $0$ & $0$ & $0$ & $0$ & $0$ & $0$ & $0$ & $0$ & $0$ & $0$ & ${\bf 100}$ & $0$ & $0$ & $0$ & $0$ & $0$ & $0$ & $0$ & $0$ & $0$ & $0$ & $0$ & $0$ \\
& $\textsc{CV}_a^\ast$ & $0$ & $9$ & ${\bf 91}$ & $0$ & $0$ & $0$ & $0$ & $0$ & $0$ & $0$ & $0$ & $0$ & $0$ & $0$ & $0$ & $0$ & $9$ & ${\bf 91}$ & $0$ & $0$ & $0$ & $0$ & $0$ & $0$ & $0$ & $0$ & $0$ & $0$ & $0$ & $0$ \\
& $\textsc{CV}_v^\ast$ & $0$ & $9$ & ${\bf 91}$ & $0$ & $0$ & $0$ & $0$ & $0$ & $0$ & $0$ & $0$ & $0$ & $0$ & $0$ & $0$ & $0$ & $9$ & ${\bf 91}$ & $0$ & $0$ & $0$ & $0$ & $0$ & $0$ & $0$ & $0$ & $0$ & $0$ & $0$ & $0$ \\ \hline
\parbox[t]{3mm}{\multirow{7}{*}{\rotatebox[origin=c]{90}{Ex-5}}}
& {\sc Dunn}$^\ast$ & $0$ & ${\bf 100}$ & $0$ & $0$ & $0$ & $0$ & $0$ & $0$ & $0$ & $0$ & $0$ & $0$ & $0$ & $0$ & $0$ & $0$ & ${\bf 100}$ & $0$ & $0$ & $0$ & $0$ & $0$ & $0$ & $0$ & $0$ & $0$ & $0$ & $0$ & $0$ & $0$ \\
& \textsc{KL}$^\ast$ & $0$ & ${\bf 100}$ & $0$ & $0$ & $0$ & $0$ & $0$ & $0$ & $0$ & $0$ & $0$ & $0$ & $0$ & $0$ & $0$ & $0$ & ${\bf 100}$ & $0$ & $0$ & $0$ & $0$ & $0$ & $0$ & $0$ & $0$ & $0$ & $0$ & $0$ & $0$ & $0$ \\
& \textsc{Gap}$^?$ & $0$ & ${\bf 19}$ & $12$ & $29$ & $28$ & $10$ & $2$ & $0$ & $0$ & $0$ & $0$ & $0$ & $0$ & $0$ & $0$ & $0$ & ${\bf 0}$ & $1$ & $17$ & $47$ & $28$ & $7$ & $0$ & $0$ & $0$ & $0$ & $0$ & $0$ & $0$ & $0$ \\
& \textsc{Jump}$^\ast$ & $0$ & ${\bf 100}$ & $0$ & $0$ & $0$ & $0$ & $0$ & $0$ & $0$ & $0$ & $0$ & $0$ & $0$ & $0$ & $0$ & $0$ & ${\bf 100}$ & $0$ & $0$ & $0$ & $0$ & $0$ & $0$ & $0$ & $0$ & $0$ & $0$ & $0$ & $0$ & $0$ \\
& $\textsc{CV}_a^\ast$ & $0$ & ${\bf 100}$ & $0$ & $0$ & $0$ & $0$ & $0$ & $0$ & $0$ & $0$ & $0$ & $0$ & $0$ & $0$ & $0$ & $0$ & ${\bf 100}$ & $0$ & $0$ & $0$ & $0$ & $0$ & $0$ & $0$ & $0$ & $0$ & $0$ & $0$ & $0$ & $0$ \\
& $\textsc{CV}_v^\ast$ & $0$ & ${\bf 100}$ & $0$ & $0$ & $0$ & $0$ & $0$ & $0$ & $0$ & $0$ & $0$ & $0$ & $0$ & $0$ & $0$ & $0$ & ${\bf 100}$ & $0$ & $0$ & $0$ & $0$ & $0$ & $0$ & $0$ & $0$ & $0$ & $0$ & $0$ & $0$ & $0$ \\ \hline
\parbox[t]{3mm}{\multirow{7}{*}{\rotatebox[origin=c]{90}{Ex-6}}}
& {\sc Dunn}$^\ast$ & $0$ & ${\bf 100}$ & $0$ & $0$ & $0$ & $0$ & $0$ & $0$ & $0$ & $0$ & $0$ & $0$ & $0$ & $0$ & $0$ & $0$ & ${\bf 100}$ & $0$ & $0$ & $0$ & $0$ & $0$ & $0$ & $0$ & $0$ & $0$ & $0$ & $0$ & $0$ & $0$ \\
& \textsc{KL}$^\ast$ & $0$ & ${\bf 100}$ & $0$ & $0$ & $0$ & $0$ & $0$ & $0$ & $0$ & $0$ & $0$ & $0$ & $0$ & $0$ & $0$ & $0$ & ${\bf 100}$ & $0$ & $0$ & $0$ & $0$ & $0$ & $0$ & $0$ & $0$ & $0$ & $0$ & $0$ & $0$ & $0$ \\
& \textsc{Gap}$^?$ & $0$ & ${\bf 100}$ & $0$ & $0$ & $0$ & $0$ & $0$ & $0$ & $0$ & $0$ & $0$ & $0$ & $0$ & $0$ & $0$ & $0$ & ${\bf 100}$ & $0$ & $0$ & $0$ & $0$ & $0$ & $0$ & $0$ & $0$ & $0$ & $0$ & $0$ & $0$ & $0$ \\
& \textsc{Jump}$^\ast$ & $0$ & ${\bf 100}$ & $0$ & $0$ & $0$ & $0$ & $0$ & $0$ & $0$ & $0$ & $0$ & $0$ & $0$ & $0$ & $0$ & $0$ & ${\bf 100}$ & $0$ & $0$ & $0$ & $0$ & $0$ & $0$ & $0$ & $0$ & $0$ & $0$ & $0$ & $0$ & $0$ \\
& $\textsc{CV}_a^\ast$ & $0$ & ${\bf 100}$ & $0$ & $0$ & $0$ & $0$ & $0$ & $0$ & $0$ & $0$ & $0$ & $0$ & $0$ & $0$ & $0$ & $0$ & ${\bf 100}$ & $0$ & $0$ & $0$ & $0$ & $0$ & $0$ & $0$ & $0$ & $0$ & $0$ & $0$ & $0$ & $0$ \\
& $\textsc{CV}_v^\ast$ & $0$ & ${\bf 100}$ & $0$ & $0$ & $0$ & $0$ & $0$ & $0$ & $0$ & $0$ & $0$ & $0$ & $0$ & $0$ & $0$ & $0$ & ${\bf 100}$ & $0$ & $0$ & $0$ & $0$ & $0$ & $0$ & $0$ & $0$ & $0$ & $0$ & $0$ & $0$ & $0$ \\ \hline
\end{tabular}
\end{center}
\vspace{-0.25in}
\begin{flushleft}
\footnotesize
~~~~~~~~~Figures in bold indicate frequencies corresponding to $k_0$.~$^\ast$ Results obtained using methods based on $\rho_0$.\\
~~~~~~~~~~~~~~~~~~~~$^?$ Both the Euclidean distance and $\rho_0$ were used, and the best result is reported
\end{flushleft}
\vspace{-0.2in}
\end{table}
Results for Examples 7 and 8 are given in Table~\ref{table::simulation_k_7&8}. Since the performance of the {\sc{Gap}} statistic was inferior to other methods, those results are not reported in Table~\ref{table::simulation_k_7&8}. In these two examples, MADD versions of different methods did not have satisfactory performance when $\rho_0$ was used, but those based on $\rho_1$ and $\rho_2$, particularly the latter ones, had much improved performance. This is consistent with what we observed in Section~\ref{sec::simulation-cluster}.

\begin{table}[h!]
\spacingset{1}
\setlength{\tabcolsep}{0.2mm}
\begin{center}
\caption{Frequency distribution for the estimated number of clusters in Examples 7 and 8}
\label{table::simulation_k_7&8}
\vspace{-0.1in}
\footnotesize
\begin{tabular}{|c|c|c|cccccccccc||cccccccccc||cccccccccc|} \hline
 & & & \multicolumn{10}{c||}{$\rho_0$} & \multicolumn{10}{c||}{$\rho_1$} & \multicolumn{10}{c|}{$\rho_2$}\\ \cline{4-33}
& & $k$  & ~1~ & ~2~ & ~3~ & ~4~ & ~5~ & ~6~ & ~7~ & ~8~ & ~9~ & 10 &~1~ & ~2~ & ~3~ & ~4~ & ~5~ & ~6~ & ~7~ & ~8~ & ~9~ & 10 & ~1~ & ~2~ & ~3~ & ~4~ & ~5~ & ~6~ & ~7~ & ~8~ & ~9~ & 10\\ \hline
\parbox[t]{3mm}{\multirow{10}{*}{\rotatebox[origin=c]{90}{Ex-7}}}
& \parbox[t]{3mm}{\multirow{5}{*}{\rotatebox[origin=c]{90}{AvgL}}}
&  {\sc Dunn}$^\ast$
& 0 & 100 & 0 & \bf 0& 0& 0& 0& 0& 0& 0
& 0 & 1 & 0 & \bf 61& 30& 6& 2& 0& 0& 0
& 0 & 0 & 0 & \bf 100& 0& 0& 0& 0& 0& 0 \\
& & PD$^\ast$
& 58 & 42 & 0 &\bf 0 & 0 & 0 & 0 & 0 & 0 & 0
& 25 & 4 & 0 &\bf 65 & 6 & 0 & 0 & 0 & 0 & 0
& 0 & 0 & 0 &\bf 100 & 0 & 0 & 0 & 0 & 0 & 0 \\
& &  \sc KL$^\ast$
& 0 & 10 & 21 & \bf 21& 26& 11& 4& 3& 2& 2
& 0 & 5 & 1 & \bf 58& 19& 4& 4& 3& 3& 3
& 0 & 0 & 0 & \bf 100& 0& 0& 0& 0& 0& 0 \\
& &  \sc Jump$^\ast$
& 100 & 0 & 0 & \bf 0& 0& 0& 0& 0& 0& 0
& 75 & 0 & 0 & \bf 13& 11& 0& 1& 0& 0& 0
& 0 & 0 & 0 & \bf 100& 0& 0& 0& 0& 0& 0 \\
& &CV$_a$$^\ast$
&0 & 1& 1 & \bf 0 & 2 & 0 & 2 & 2 & 3 & 89
&0 & 1 & 0 & \bf 40 & 55 & 4 &0 & 0 & 0 & 0
&0 & 0 & 0 & \bf 100 & 0 & 0 &0 & 0 & 0 & 0 \\
& & CV$_v$$^\ast$
&0 & 38& 5 & \bf 10 & 10 & 8& 2 & 1 & 0 & 26
&0 & 13 & 0 & \bf 50 & 36 & 1 &0 & 0 & 0 & 0
&0 & 0 & 0 & \bf 100 & 0 & 0 &0 & 0 & 0 & 0 \\ \cline{2-33}
& \parbox[t]{3mm}{\multirow{5}{*}{\rotatebox[origin=c]{90}{kM}}}
&  {\sc Dunn}$^\ast$
&0 & 99 & 1 & \bf 0 & 0 &0 & 0 & 0 & 0 & 0
&0 & 9 & 1& \bf 86 & 4 & 0 &0 & 0 & 0 & 0
&0 & 0 & 0 & \bf 100 & 0 & 0 &0 & 0 & 0 & 0 \\
& & PD$^\ast$
& 0 & 100 & 0 &\bf 0 & 0 & 0 & 0 & 0 & 0 & 0
& 2 & 11 & 0 &\bf 86 & 1 & 0 & 0 & 0 & 0 & 0
& 0 & 0 & 0 &\bf 100 & 0 & 0 & 0 & 0 & 0 & 0 \\
& &  \sc KL$^\ast$
& 0 & 60 & 14 & \bf 4& 7& 6& 2& 3 & 2& 2
& 0 & 0 & 5 & \bf 63& 6& 2& 1& 6& 9& 8
& 0 & 0 & 0 & \bf 100& 0& 0& 0& 0& 0& 0 \\
& &  \sc Jump$^\ast$
& 100 & 0 & 0 & \bf 0& 0& 0& 0& 0& 0& 0
& 78 & 0 & 0 & \bf 15& 6& 1& 0& 0& 0& 0
& 0 & 0 & 0 & \bf 100& 0& 0& 0& 0& 0& 0 \\
& &CV$_a$$^\ast$
&0 & 32& 8 & \bf 1 & 2 & 3 & 2 & 5 & 1 & 46
&0 & 2 & 0 & \bf 69 & 29 & 0 &0 & 0 & 0 & 0
&0 & 0 & 0 & \bf 100 & 0 & 0 &0 & 0 & 0 & 0 \\
& & CV$_v$$^\ast$
&0 & 74& 21 & \bf 2 & 3 & 0& 0 & 0 & 0 & 0
&0 & 18 & 1 & \bf 79 & 2 & 0 &0 & 0 & 0 & 0
&0 & 0 & 0 & \bf 100 & 0 & 0 &0 & 0 & 0 & 0 \\ \hline \hline
\parbox[t]{3mm}{\multirow{10}{*}{\rotatebox[origin=c]{90}{Ex-8}}}
& \parbox[t]{3mm}{\multirow{5}{*}{\rotatebox[origin=c]{90}{AvgL}}}
&  {\sc Dunn}$^\ast$ &
0 & \bf 85 & 14 & 1 & 0 & 0 & 0 & 0 & 0 & 0 & 0 & \bf 100 & 0 & 0 & 0 & 0 & 0 & 0 & 0 & 0 &0 & \bf 100 & 0 & 0 & 0 & 0 & 0 & 0 & 0 & 0 \\
& & PD$^\ast$
& 37 &\bf 63 & 0 & 0 & 0 & 0 & 0 & 0 & 0 & 0
& 1 &\bf 99 & 0 & 0 & 0 & 0 & 0 & 0 & 0 & 0
& 0 &\bf 100 & 0 & 0 & 0 & 0 & 0 & 0 & 0 & 0 \\
& & \textsc{KL}$^\ast$ & 0 & \bf 18 & 17 & 14 & 22 & 14 & 8 & 5 & 1 & 1
& 0 & \bf 85 & 12 & 1 & 0 & 1 & 1 & 0 & 0 & 0   & 0 & \bf 100 & 0 & 0 & 0 & 0 & 0 & 0 & 0 & 0 \\
& & \textsc{Jump}$^\ast$ & 0 & \bf 0 & 16 & 26 & 32 & 13 & 6 & 4 & 2 & 1
& 0 & \bf 94 & 6 & 0 & 0& 0& 0& 0& 0& 0   & 0 & \bf 100 & 0 & 0 & 0& 0& 0& 0& 0& 0  \\
& & CV$_a$$^\ast$ & 0 & \bf 32 & 0 & 0 & 0 & 0 & 1 & 3 & 4 & 60
& 0 & \bf 93 & 2 & 0 & 0 & 1 & 0 & 0 & 3 & 1
& 0 & \bf 100 & 0 & 0 & 0 & 0 & 0 & 0 & 0 & 0 \\
& & CV$_v$$^\ast$ & 0 & \bf 94 & 2 & 0 & 0 & 0 & 0 & 0 & 0 & 4
& 0 & \bf 99 & 1 & 0 & 0 & 0 & 0 & 0 & 0 & 0
&0 & \bf 100 & 0 & 0 & 0 & 0 & 0 & 0 & 0 & 0 \\ \cline{2-33}
& \parbox[t]{3mm}{\multirow{5}{*}{\rotatebox[origin=c]{90}{kM}}}&
{\sc Dunn}$^\ast$
& 0 & \bf 78 & 16 & 6 & 0 & 0 & 0 & 0 & 0 &0
& 0 & \bf 99 & 1 & 0 & 0 & 0 & 0 & 0 & 0 & 0
& 0 & \bf 100 & 0 & 0 & 0 & 0 & 0 & 0 & 0 & 0 \\
& & PD$^\ast$
& 57 &\bf 42 & 1 & 0 & 0 & 0 & 0 & 0 & 0 & 0
& 0 &\bf 100 & 0 & 0 & 0 & 0 & 0 & 0 & 0 & 0
& 0 &\bf 100 & 0 & 0 & 0 & 0 & 0 & 0 & 0 & 0 \\
& & \textsc{KL}$^\ast$
& 0 & \bf 22 & 25 & 20 & 10 & 13 & 6 & 2 & 1 & 1
& 0 & \bf 96 & 4 & 0 & 0 & 0 & 0 & 0 & 0 & 0
& 0 & \bf 100 & 0 & 0 & 0 & 0 & 0 & 0 & 0 & 0 \\
& & \textsc{Jump}$^\ast$
& 1 & \bf 2 & 44 & 33 & 14 & 5 & 1 & 0 & 0 & 0
& 0 & \bf 96 & 4 & 0 & 0 & 0 & 0 & 0 & 0 & 0
& 0 & \bf 100 & 0 & 0 & 0 & 0 & 0 & 0 & 0 & 0 \\
& & CV$_a$$^\ast$
& 0 & \bf 15 & 0 & 0 & 0 & 1 & 2 & 2 & 2 & 78
& 0 & \bf 95 & 1 & 0 & 0 & 0 & 0 & 0 & 1 & 3
& 0 & \bf 100 & 0 & 0 & 0 & 0 & 0 & 0 & 0 & 0 \\
& & CV$_v$$^\ast$
& 0 & \bf 83 & 3 & 0 & 0 & 3 & 0 & 2 & 5 & 4
& 0 & \bf 100 & 0 & 0 & 0 & 0 & 0 & 0 & 0 & 0& 0 & \bf 100 & 0 & 0 & 0 & 0 & 0 & 0 & 0 & 0 \\ \hline
\end{tabular}
\end{center}
\vspace{-0.3in}
\begin{flushleft}
\footnotesize
~~~~~~~~~~Figures in bold indicate frequencies corresponding to $k_0$.~$^\ast$ Results are obtained using MADD versions.
\end{flushleft}
\vspace{-0.25in}
\end{table}

The success of KL statistic, {\sc{Jump}} statistic and {\sc Dunn} index in all examples motivated us to carry out a theoretical investigation regarding their high dimensional behavior. For this investigation, we make some assumptions on  asymptotic orders of $\rho_{h,\psi}$. For two independent observations $\Xvec$ and $\Yvec$ from $i$-th and $j$-th populations, let $\rho_{h,\psi}(\Xvec,\Yvec) \overset{P}{\asymp} \phi_{ij}(d)$, 
i.e.,  as $d \to \infty$, $\Pr\big(\rho_{h,\psi}(\Xvec,\Yvec)/\phi_{ij}(d) \text{ remains bounded away from } 0 \text{ and } \infty\big) \to 1$. Here we assume that
\begin{enumerate}[$({A}1)$]
\vspace{-0.1in}
\setcounter{enumi}{2}
\item {\it $\phi_{ii}(d) \asymp \phi_{-}(d)$ for every $i=1,\ldots,k_0$ and $\phi_{ij}(d) \asymp \phi_{+}(d)$ for every $1 \le i \ne j \le k_0$, where $\phi_{-}(d)= {\bf o}(\phi_{+}(d))$.}
\vspace{-0.11in}
\end{enumerate}
If $\Xvec$ and $\Yvec$ come from the same population, under $(A1)$ we have $\rho_{h,\psi}(\Xvec,\Yvec)={\bf o}_P(1)$. So, $\phi_{-}(d)$ should decrease to $0$ as $d$ increases. It also follows from $(A1)$ and $(A2)$ that if $\Xvec$ and $\Yvec$ are from the $i$-th and the $j$-th populations $(i \neq j)$, $|\rho_{h,\psi}(\Xvec,\Yvec)-\rho^{*}_{h,\psi}(i,j)| \stackrel{P}{\rightarrow}0$, where $\liminf_{d \to \infty} \rho^{*}_{h,\psi}(i,j) > 0$. So, $\phi_{+}(d)$ remains bounded away from $0$ as $d$ increases. Thus, $(A3)$ holds trivially under $(A1)$ and $(A2)$. Note that under $(A2^\circ)$ also, we have $\phi_{-}(d)/\phi_{+}(d)= {\bf O}(\vartheta(d)/d\;\rho_{h,\psi}^\ast(i,j))={\bf o}(1)$.

Under this assumption, MADD versions of \textsc{KL} statistic, \textsc{Jump} statistic and \textsc{Dunn} index have some nice properties in high dimensions if an appropriate base clustering algorithm is used. To make it clear what we mean by an appropriate base clustering algorithm, we now introduce the concept of perfect and order preserving (POP) clustering.

\begin{defn}\label{def::POP}
For any fixed $k$, let $C_1^{(k)},\ldots,C_k^{(k)}$ be $k$ clusters estimated using a clustering algorithm on ${\cal X} = \cup_{i=1}^{k_0} {\cal X}_i$, which consists of observations from $k_0$ classes. We call the algorithm perfect and order preserving (POP) at $k_0$ if the following conditions hold.
\vspace{-0.05in}
\begin{enumerate}[$(a)$]
\spacingset{0.75}
\item The clustering algorithm is perfect, i.e., for $k=k_0$, $C_i^{(k_0)} = {\cal X}_{\pi(i)}$ for every $i=1,\ldots,k_0$ and some permutation $\pi$ of $\{1,\ldots,k_0\}$.
\vspace{-0.025in}
\item For any $k<k_0$ and for every $i=1,\ldots,k_0$, there exists $j \le k$ such that $C_i^{(k_0)} \subseteq C_j^{(k)}$.
\vspace{-0.025in}
\item For any $k>k_0$ and for every $i=1,\ldots,k$, there exits $j \le k_0$ such that $C_i^{(k)} \subseteq C_j^{(k_0)}$.
\end{enumerate}
\end{defn}

\vspace{-0.15in}
\begin{figure}[h]
\centering
\includegraphics[height=2.20in,width=3.0in]{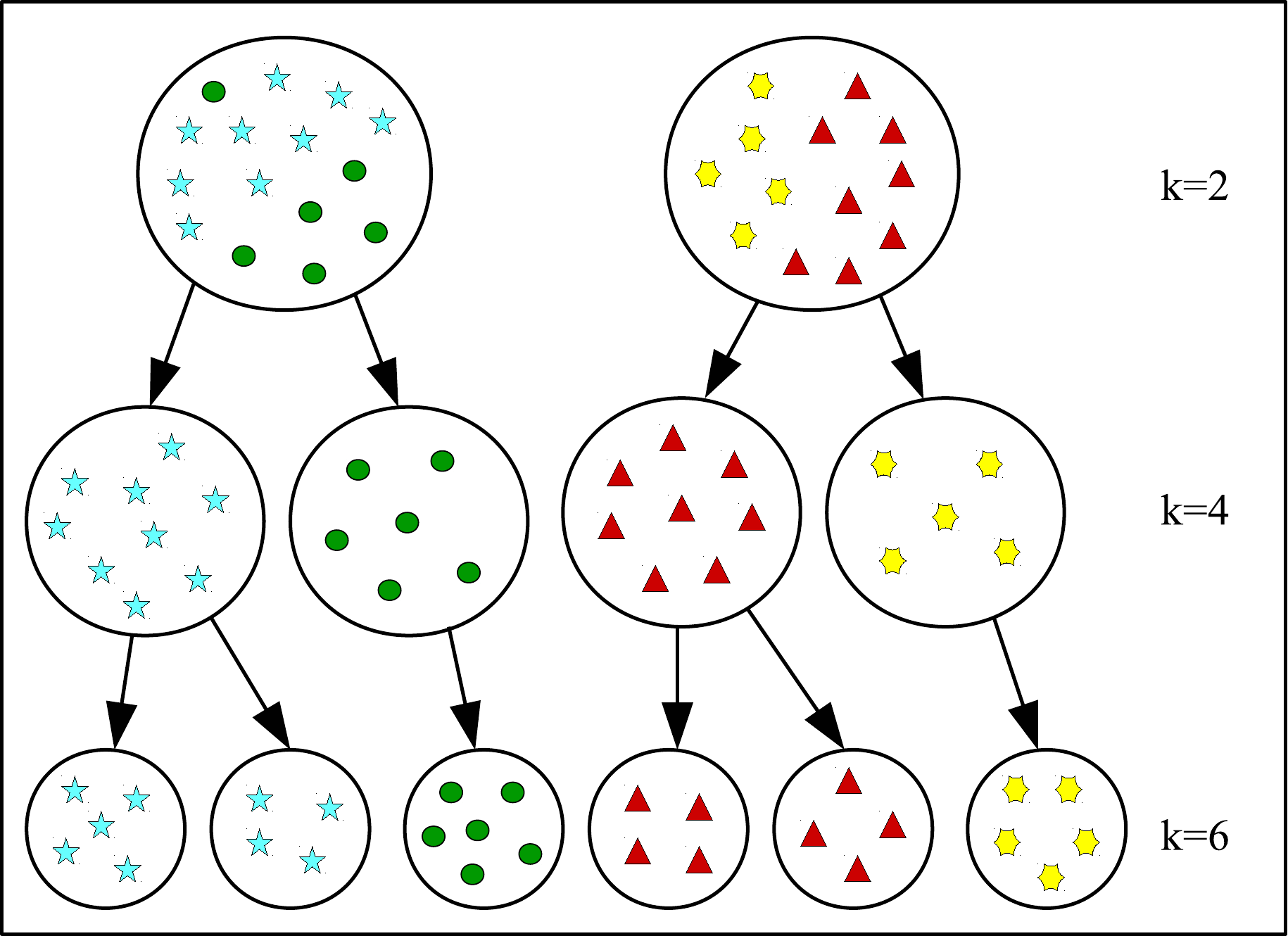}

\vspace{-0.15in}
\caption{A clustering algorithm which is POP at $4$.}
\label{fig:POP}
\vspace{-0.1in}
\end{figure}

Figure~\ref{fig:POP} demonstrates a POP clustering (at $4$) by taking only one value of $k$ smaller than $4$ and one value of $k$ bigger than $4$. But, one should notice that property (b) (respectively, property (c)) has to hold for all $k<4$ (respectively, $k>4$). It is easy to check that any hierarchical algorithm is order preserving (i.e., satisfies (b) and (c)). So, if it leads to perfect clustering (i.e., satisfies (a)), it becomes POP at $k_0$. In Theorems~\ref{thm::exact cluster hierarchical}--\ref{thm::MADD_Euclid_CLT}, we have seen that AvgL($h,\psi$) and kM($h, \psi$) become perfect with probability tending to one as $d$ tends to infinity. Using Lemmas~\ref{lemma:MADD_asymptotic}--\ref{lemma:MADD_euclidasymptotic}, one can show that, they become POP at $k_0$ with probability tending to one as the dimension diverges. Assuming that such a POP algorithm is used for base clustering, the following theorem shows the high dimensional behavior of estimators based on the MADD versions of {\sc Dunn} index, \textsc{KL} statistic and \textsc{Jump} statistic.

\vspace{-0.05in}
\begin{theorem}\label{thm::consistency Dunn}
Suppose that there are observations from $k_0 \ge 2$ populations which satisfy $(A3)$, and also assume that the base clustering algorithm is POP at $k_0$.\\Then
$(i)$ $\hat{k}_D^* \overset{P}{\rightarrow} k_0$, $(ii)$ $\hat{k}_{KL}^* \overset{P}{\rightarrow} k_0$ and $(iii)$ $\Pr(\hat{k}_{J}^* \ge k_0) \to 1$ as $d \to \infty$.\\
$($Here $\hat{k}_D^*$, $\hat{k}_{KL}^*$ and $\hat{k}_{J}^*$ are the number of clusters estimated by MADD versions of {\sc Dunn} index, \textsc{KL} statistic and \textsc{Jump} statistic $($with $t=1)$, respectively.$)$
\end{theorem}
\vspace{-0.1in}

Theorem~\ref{thm::consistency Dunn} shows the high dimensional consistency of ${\hat k}^*_{KL}$ and ${\hat k}^*_{D}$. But since $\textsc{D}(1)$ and $\textsc{KL}(1)$ are not defined, they cannot detect the presence of a single cluster. \textsc{Jump} statistic can be used in such situations, but Theorem~\ref{thm::consistency Dunn} only shows that $\Pr(\hat{k}_{J}^* \ge k_0) \to 1$ as $d \to \infty$. So, it can overestimate $k_0$ in some cases. To overcome these limitations, we define a penalized version of {\sc Dunn} index (PD). For any fixed $k$, it is given by $\textsc{PD}(k) = B_k^{\circ}/W_k^{\circ} - k \zeta(d)$, where $B_k^{\circ}$ (for $k \ge 2$) and $W_k^{\circ}$ have the same meaning as in the {\sc Dunn} index, $B_1^{\circ}\overset{def}{=} B_2^{\circ}$ and $\zeta$ is the penalty function. We estimate $k_0$ by maximizing $\textsc{PD}(k)$ with respect to $k$ and denote it by $\hat{k}_{PD}$ ($\hat{k}^*_{PD}$ when MADD versions are used). The following theorem shows the high dimensional consistency of $\hat{k}^*_{PD}$ for suitable choices of $\zeta$.

\begin{theorem}\label{thm::consistency penalized Dunn}
Suppose that there are observations from $k_0 \ge 1$ population(s), which satisfy $(A3)$, and the penalty function $\zeta(d)\rightarrow \infty$ in such a way that $\phi_{-}(d)\zeta(d)/\phi_{+}(d) \rightarrow 0$
as $d\rightarrow \infty$. If the base clustering algorithm is POP at $k_0$, then $\hat{k}^*_{PD} \overset{P}{\rightarrow} k_0$ as $d \to \infty$.
\end{theorem}
\vspace{-0.13in}

We have already seen that under $(A1)$ and $(A2)$, while $\phi_{+}(d)$ remains bounded away from $0$, $\phi_{-}(d)$ converges to $0$ as $d$ increases. So, if we assume $\phi_{-}(d)={\bf O}(d^{-\alpha_0})$ for some $\alpha_0>0$, one can use any $\zeta$ such that $1/\zeta(d)$ decreases to zero at a slower rate than ${\bf O}(d^{-\alpha_0})$. For instance, one can use $\zeta(d) = \lambda \log(d)$ for a suitable choice of the parameter $\lambda$. Some sufficient conditions for $\phi_{-}(d)={\bf O}(d^{-\alpha_0})$ are given in the Appendix (see Lemma~\ref{lemma:MADD_sameclass} and the remark after the proof of Lemma~\ref{lemma:MADD_sameclass}). Throughout this article, we used $\zeta(d) =\lambda \log(d)$, where $\lambda=0.015$ was chosen based on our empirical experience. This choice of $\zeta$ worked well in all simulated and real data sets we analyzed in this article.

In Examples 1--6, $\hat{k}^*_{PD}$ had same results as obtained using $\hat{k}^*_D$. Use of $\rho_0$, $\rho_1$ and $\rho_2$  yielded similar results in these examples. In Examples 7 and 8, however, the MADD version of \textsc{PD} did not have satisfactory results when $\rho_0$ was used. In many cases, it failed to identify the underlying clusters, and $\hat{k}^*_{PD}$ turned out to be $1$. Using $\rho_1$ and $\rho_2$, we got better results in these two examples (see the results corresponding to {\sc PD}$^\ast$ in Table~\ref{table::simulation_k_7&8}). Among these two choices, the latter one yielded better results. For further evaluation of the performance of ${\hat k}^*_{PD}$, we generated $100$ observations from a uniform distribution on the $500$-dimensional unit hypercube, and repeated the experiment $100$ times. In all these $100$ cases, it successfully identified the presence of a single cluster in the data set for all three choices of $\rho_{h,\psi}$.

Note that Theorems~\ref{thm::consistency Dunn} and \ref{thm::consistency penalized Dunn} show the consistency of $\hat{k}^*_{KL}$, $\hat{k}^*_{D}$ and $\hat{k}^*_{PD}$ when all within cluster separations are of the same asymptotic order and so are the between cluster separations, i.e., $\phi_{ii}(d) \asymp \phi_{-}(d)$ for all $i$ and $\phi_{ij}(d) \asymp \phi_{+}(d)$ for all $i \ne j$. If that is not the case but $\max_i \phi_{ii}(d)={\bf o}(\min_{i \neq j} \phi_{ij}(d))$, these methods may detect $k_0^\prime (< k_0)$ super-clusters in the data, each consisting of one or more clusters (can be proved using similar arguments as used in the proofs of Theorems~\ref{thm::consistency Dunn} and \ref{thm::consistency penalized Dunn}). In that case, instead of stopping after one step, we need to repeat the algorithm on each of the estimated super-clusters. One can use the penalized {\sc Dunn} index (with appropriate penalty function) for this purpose and stop splitting a super-cluster when $\hat{k}^\ast_{PD}$ turns out to be $1$. One can check that this repetitive use of {\sc PD} consistently estimates $k_0$. However, we did not use this repetitive method in this article.

\vspace{-0.15in}
\section{Analysis of benchmark data sets}\label{sec::benchmark}

\vspace{-0.05in}
We analyzed two benchmark data sets, `Lymphoma' data and `Control Chart' data, for further evaluation of our proposed methods. Lymphoma data set was first analyzed by \cite{Lymph00} for identification of distinct types of lymphoma, and it is available in the \texttt{R} package \texttt{spls}. Control Chart data set can be obtained from the \texttt{UCI Machine Learning Repository} (\url{https://archive.ics.uci.edu/ml/datasets.html}).

\vspace{-0.2in}
\subsection{Lymphoma data}

\vspace{-0.1in}
This data set contains expression levels of $4026$ genes for $42$ diffuse large B-cell lymphoma ({\tt DLBCL}), $9$ follicular lymphoma ({\tt FL}) and $11$ chronic lymphocytic leukemia ({\tt CLL}) samples. A plot of these $62$ observations is given in Figure~\ref{fig::lymphoma}.

\definecolor{DLBCL}{rgb}{0,1,0}
\definecolor{FL}{rgb}{0,0,0.9}
\definecolor{CLL}{rgb}{1,0.3,0.3}
\begin{figure}[h!]
\centering
\includegraphics[width=6.0in, height=1.80in]{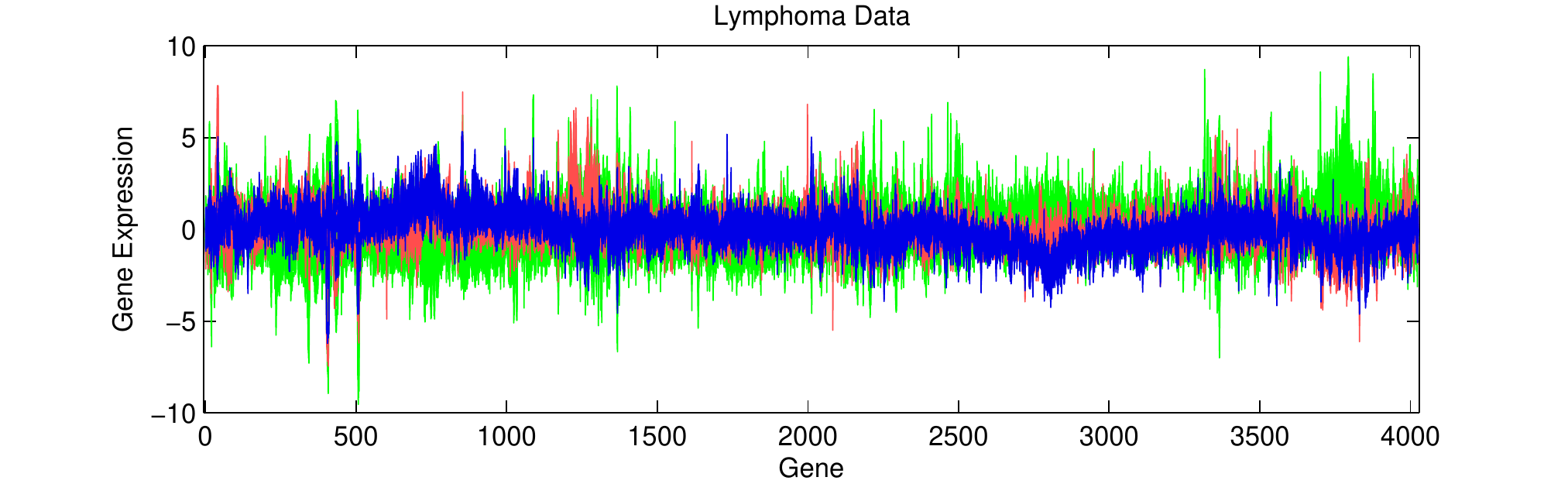}

\vspace{-0.2in}
\caption{Gene expression of $4026$ genes for {\tt DLBCL}(\crule[DLBCL]{0.1in}{0.1in}), {\tt FL}(\crule[FL]{0.1in}{0.1in}) and {\tt CLL}(\crule[CLL]{0.1in}{0.1in}).}
\label{fig::lymphoma}
\vspace{-0.1in}
\end{figure}

\begin{figure}[h!]
\spacingset{1}
\centering
\begin{subfigure}[b]{\linewidth}
\centering
\subcaption{Compositions of 2 clusters}
\vspace{-0.1in}
\includegraphics[width=5.5in,height=1.25in]{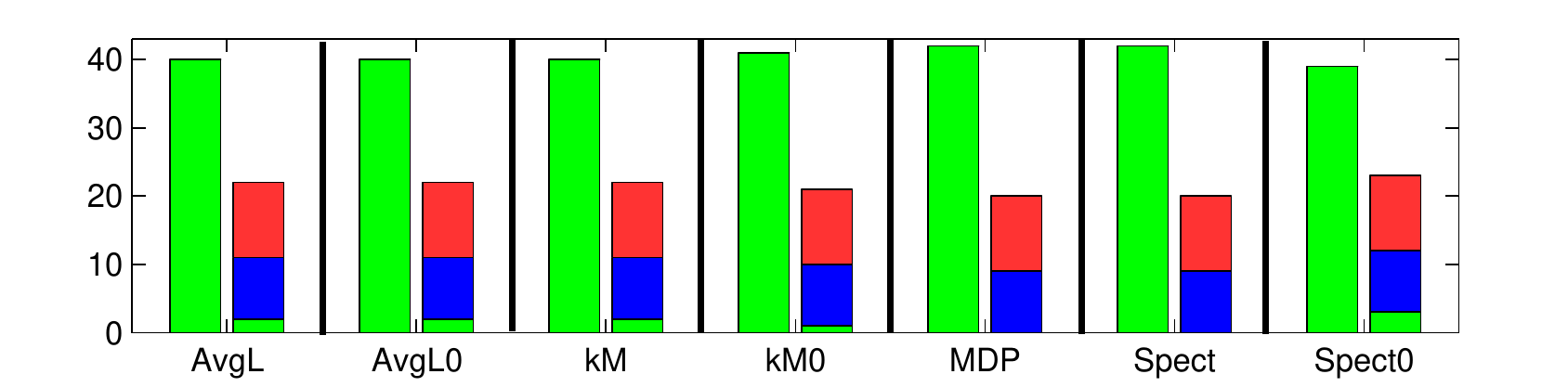}
\end{subfigure}
\vspace{-0.05in}
\begin{subfigure}[b]{\linewidth}
\centering
\subcaption{Compositions of 3 clusters}
\vspace{-0.1in}
\includegraphics[width=5.5in,height=1.25in]{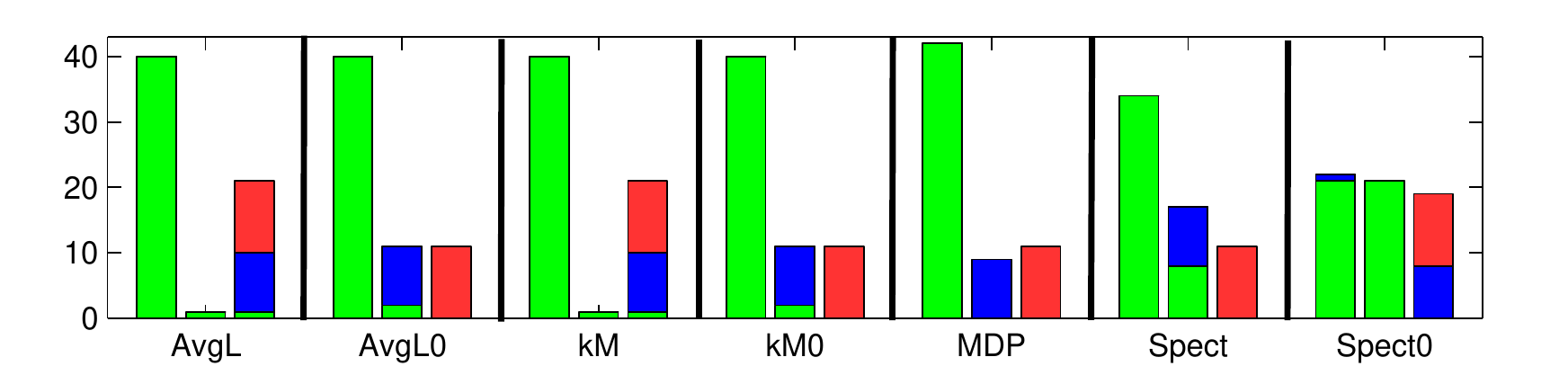}
\end{subfigure}

\vspace{-0.1in}
\caption{\small Compositions of (a) two and (b) three estimated clusters for Lymphoma data. Each bar corresponds to a single cluster consisting of {\tt DLBCL}(\crule[DLBCL]{0.1in}{0.1in}), {\tt FL}(\crule[FL]{0.1in}{0.1in}) and {\tt CLL}(\crule[CLL]{0.1in}{0.1in}) samples}
\label{fig::lymphoma_clusters}
\vspace{-0.1in}
\end{figure}

We used different methods to estimate $k_0$ and the results are given in Table~\ref{table::lymphoma_k}. This table shows that all methods, except the \textsc{Gap} statistic, identified two clusters. When we used different clustering algorithms to estimate these two clusters, all of them put almost all {\tt DLBCL} samples in one cluster and the rest in another cluster (see Figure~\ref{fig::lymphoma_clusters}(a)). This indicates that it is very hard to distinguish between {\tt FL} and {\tt CLL} samples, which can be seen in Figure~\ref{fig::lymphoma} as well. This claim is also justified by the behavior of {\tt FL}, which can sometimes present itself as {\tt CLL} (see \sloppy\url{https://en.wikipedia.org/wiki/B-cell_chronic_lymphocytic_leukemia}).
\begin{table}[h]
\centering
\spacingset{1}
\caption{Number of clusters estimated by different methods in `Lymphoma' data}
\vspace{-0.1in}
\label{table::lymphoma_k}
\begin{tabular}{|c|c|c|c|c|c|c|c|} \hline
 & {\sc Dunn} & \textsc{PD} & \textsc{KL} & \textsc{Gap} & \textsc{Jump} & $\textsc{CV}_a$ & $\textsc{CV}_v$ \\ \hline
AvgL & $2$ & $2$ & $2$ & $12$ & $2$ & $2$ & $2$ \\ \hline
AvgL$_0$ & $2$ & $2$ & $2$ & $7$ & $2$ & $2$ & $2$ \\ \hline
kM & $2$ & $2$ & $2$ & $12$ & $2$ & $2$ & $2$ \\ \hline
kM$_0$ & $2$ & $2$ & $2$ & $7$ & $2$ & $2$ & $2$ \\ \hline
\end{tabular}
\vspace{-0.15in}
\end{table}

Since it was known that the observations were actually from three populations, we used different clustering algorithms to find three clusters in this data set as well. In Figure~\ref{fig::lymphoma_clusters}(b), one can see that both AvgL and kM failed to identify the three populations. But, AvgL$_0$ and kM$_0$ successfully differentiated between observations from {\tt FL} and {\tt CLL} classes. The method based on MDP led to perfect clustering, but spectral clustering algorithms did not perform well. Since all three choices of $\rho_{h, \psi}$ (i.e., $\rho_0$, $\rho_1$ and $\rho_2$) led to similar results in this data set, here we have reported the results for MADD versions based on $\rho_0$ only.

\vspace{-0.2in}
\subsection{Control chart data}

\begin{figure}[h!]
\centering
\includegraphics[width=6.5in,height=2.2in]{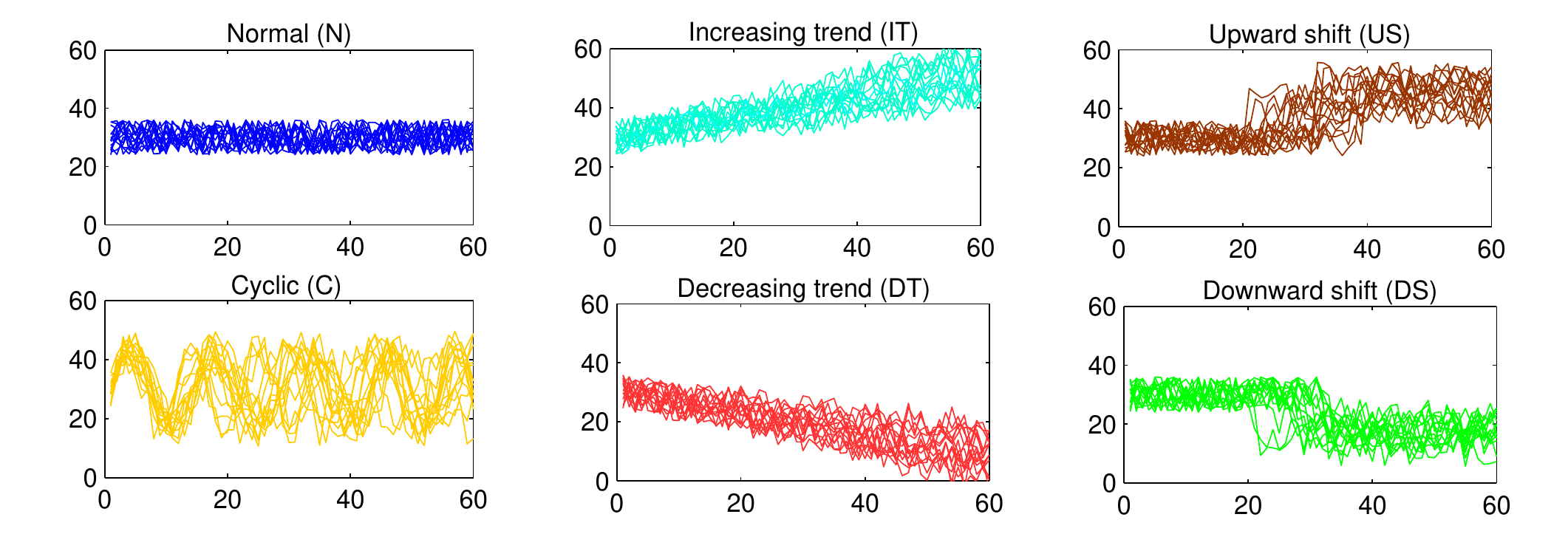}

\vspace{-0.2in}
\caption{Six classes in `Control Chart' data}
\label{fig::Synthetic}
\vspace{-0.1in}
\end{figure}

\vspace{-0.1in}
This data set contains $60$ dimensional observations from $6$ classes, viz., normal({\tt N}), cyclic({\tt C}), increasing trend({\tt IT}), decreasing trend({\tt DT}), upward shift({\tt US}) and downward shift({\tt DS}). We have 100 observations from each class. Figure~\ref{fig::Synthetic} depicts a representation of the $6$ classes.

{\sc Dunn} index and \textsc{PD} could find only two clusters in this data set, but most of their MADD versions identified three or more clusters, as did most other methods (see Table~\ref{table::synthetic_k}).  \textsc{Gap} statistic again overestimated $k_0$. \textsc{Jump} statistic also overestimated $k_0$ in some cases, but when $\rho_2$ was used, $\hat{k}^*_J$ turned out to be $1$. It also turned out to be $6$ in some cases, but those estimated clusters did not correspond to the six classes, as one can see from Figure~\ref{fig::synthetic_cluster_k=4.6}.

\begin{table}[h!]
\spacingset{1}
\centering
\caption{Number of clusters estimated by different methods in `Control Chart' data}
\label{table::synthetic_k}
\begin{tabular}{|c|c|c|c|c|c|c|c|} \hline
 & {\sc Dunn} & \textsc{PD} & \textsc{KL} & \textsc{Gap} & \textsc{Jump} & $\textsc{CV}_a$ & $\textsc{CV}_v$ \\ \hline
AvgL & $2$ & $2$ & $3$ & $10$ & $8$ & $3$ & $3$ \\
AvgL$_0$ & $3$ & $3$ & $10$ & $8$ & $6$ & $3$ & $3$ \\
AvgL$_1$ & $3$ & $3$ & $10$ & $10$ & $10$ & $3$ & $3$ \\
AvgL$_2$ & $3$ & $2$ & $11$ & $7$ & $1$ & $4$ & $4$ \\ \hline
kM & $2$ & $2$ & $3$ & $9$ & $3$ & $3$ & $3$ \\
kM$_0$ & $3$ & $3$ & $3$ & $10$ & $10$ & $3$ & $2$ \\
kM$_1$ & $3$ & $3$ & $3$ & $10$ & $6$ & $3$ & $3$ \\
kM$_2$ & $4$ & $2$ & $6$ & $7$ & $1$ & $4$ & $4$ \\ \hline

\end{tabular}
\vspace{-0.1in}
\end{table}

Since most of the methods identified two or three clusters in this data set, at first we used different clustering algorithms for finding those two or  three clusters. These results are shown in Figure~\ref{fig::synthetic_cluster_k=2.3}. MDP clustering had poor performance in this example. Since the dimension was smaller than the sample size, it was quite expected in view of the results reported in Figure~\ref{fig:examplesA&B} and Tables~\ref{table::simulation rand}--\ref{table::Rand_7+8}. So, results for MDP clustering are not reported here. Results for $\rho_0$ and $\rho_1$ were almost similar, but those for $\rho_2$ were somewhat different. So, we reported the results based on $\rho_0$ and $\rho_2$ only.

Figure~\ref{fig::synthetic_cluster_k=2.3}(a) shows that when different clustering algorithms were used to divide the data set into two groups, most of them put the observations from classes {\tt IT} and {\tt US} in one cluster and the rest in the other cluster. Methods based on $\rho_2$ led to different cluster formations. Spect put the observations from the class {\tt IT} in a cluster and the rest in  another cluster.

\definecolor{normal}{rgb}{0,0,1}
\definecolor{cyclic}{rgb}{1,0.8,0}
\definecolor{inc}{rgb}{0,1,0.8}
\definecolor{dec}{rgb}{1,0.2,0.2}
\definecolor{up}{rgb}{0.6,0.2,0}
\definecolor{down}{rgb}{0,1,0}
\vspace{-0.1in}
\begin{figure}[h!]
\centering
\begin{subfigure}[b]{\linewidth}
\centering
\subcaption{Compositions of 2 clusters}
\vspace{-0.1in}
\includegraphics[width=5.5in,height=1.25in]{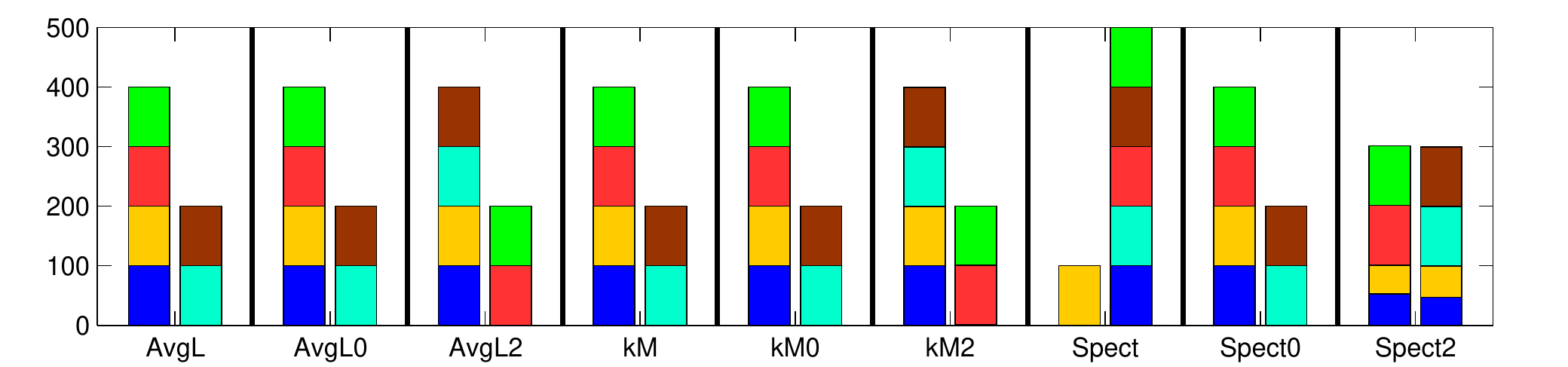}
\vspace{-0.05in}
\end{subfigure}
\begin{subfigure}[b]{\linewidth}
\centering
\subcaption{Compositions of 3 clusters}
\vspace{-0.1in}
\includegraphics[width=6.0in,height=1.25in]{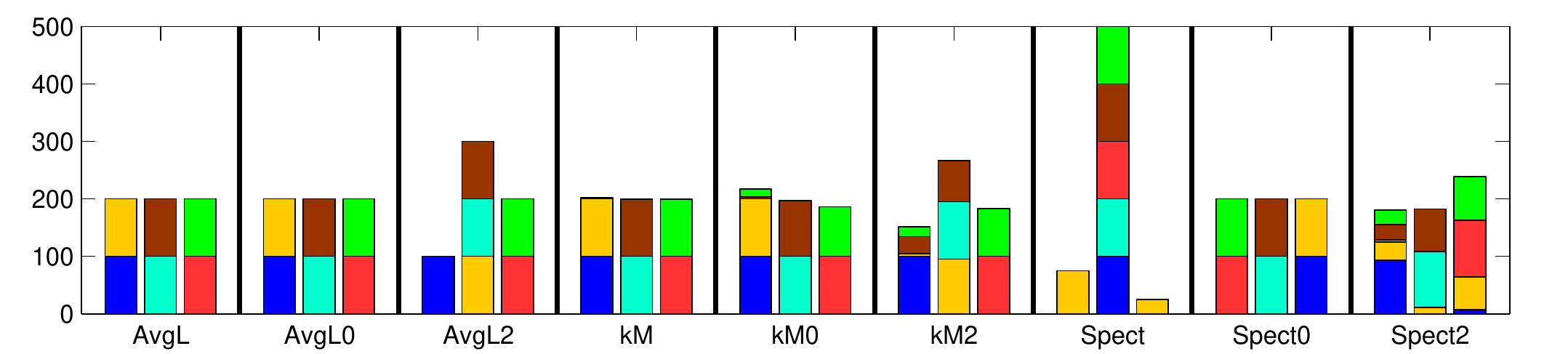}
\end{subfigure}

\vspace{-0.2in}
\caption{\small Compositions of (a) two and (b) three estimated clusters. Each bar corresponds to a single cluster, comprising of observations from different classes (normal(\crule[normal]{0.1in}{0.1in}), cyclic(\crule[cyclic]{0.1in}{0.1in}), increasing trend(\crule[inc]{0.1in}{0.1in}), decreasing trend(\crule[dec]{0.1in}{0.1in}), upward shift(\crule[up]{0.1in}{0.1in}) and downward shift(\crule[down]{0.1in}{0.1in})).}
\label{fig::synthetic_cluster_k=2.3}
\vspace{-0.1in}
\end{figure}

When we divided the data set into three clusters, AvgL, AvgL$_0$, kM and kM$_0$ formed one cluster mainly consisting of {\tt N} and {\tt C} samples; one cluster mainly consisting of {\tt DT} and {\tt DS} samples, while the third cluster was formed mainly by {\tt IT} and {\tt US} samples as before (see Figure~\ref{fig::synthetic_cluster_k=2.3}(b)). Again, $\rho_2$ led to slightly different formation of clusters. Performance of AvgL and AvgL$_0$ was slightly better than kM and kM$_0$. Spect performed poorly, but Spect$_0$ performed much better. It led to the same clusters as obtained by AvgL and AvgL$_0$.

Figure~\ref{fig::synthetic_cluster_k=4.6}(a) shows the clusters estimated by different methods when the observations were divided into four clusters. In this case, AvgL and kM divided the cluster containing {\tt N} and {\tt C} samples to form two new clusters, one containing half of the {\tt C} samples, and the other containing the rest. However, AvgL$_2$ and kM$_2$ successfully separated {\tt N} and {\tt C} samples. Performances of Spect and Spect$_0$ were similar, but Spect$_2$ yielded different results.

\begin{figure}[h!]
\vspace{-0.1in}
\centering
\begin{subfigure}[b]{\linewidth}
\centering
\subcaption{Compositions of 4 clusters}
\vspace{-0.1in}
\includegraphics[width=5.0in,height=2.5in]{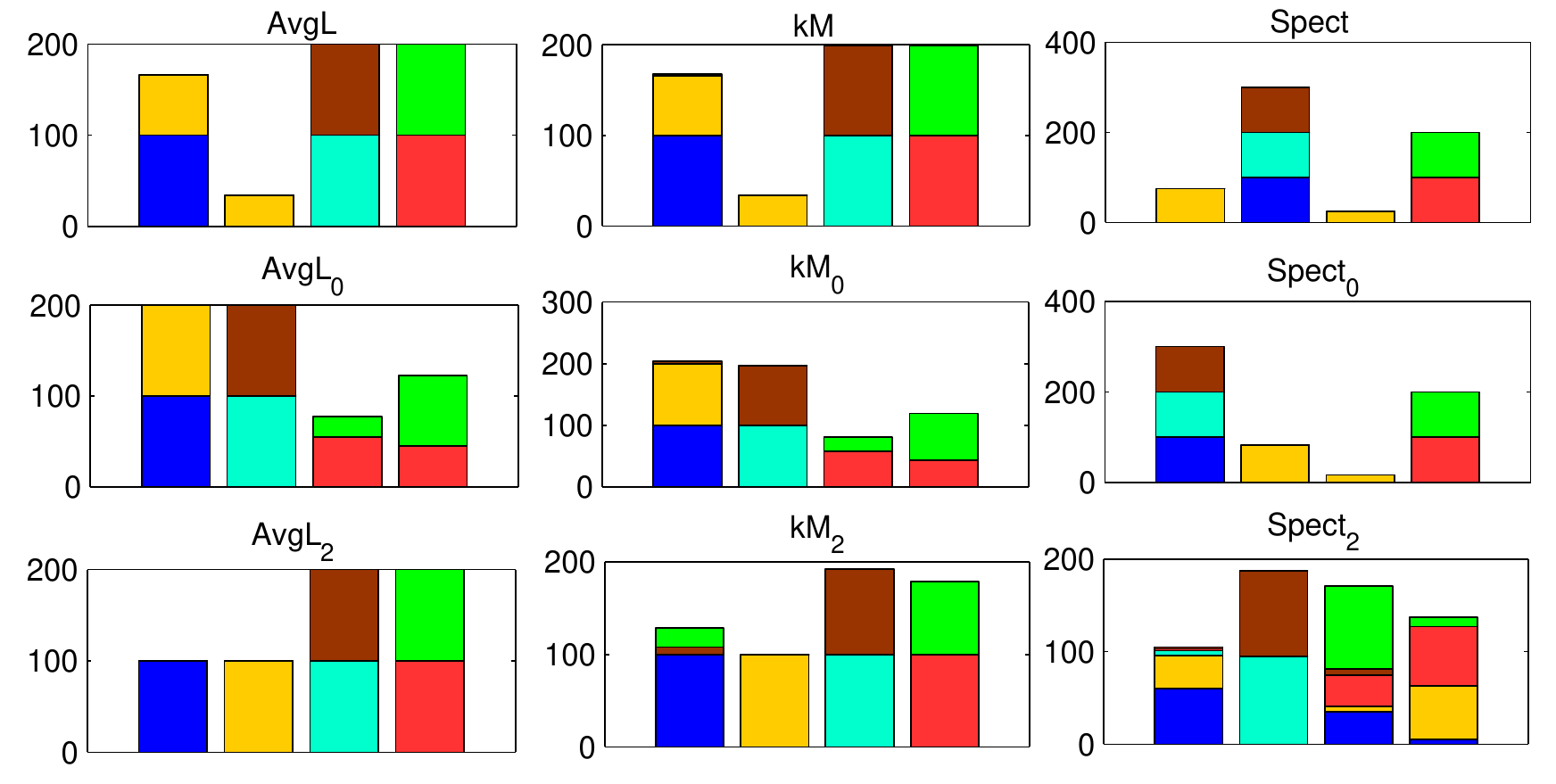}
\vspace{-0.1in}
\end{subfigure}

\begin{subfigure}[b]{\linewidth}
\centering
\subcaption{Compositions of 6 clusters}
\vspace{-0.1in}
\includegraphics[width=5.5in,height=2.50in]{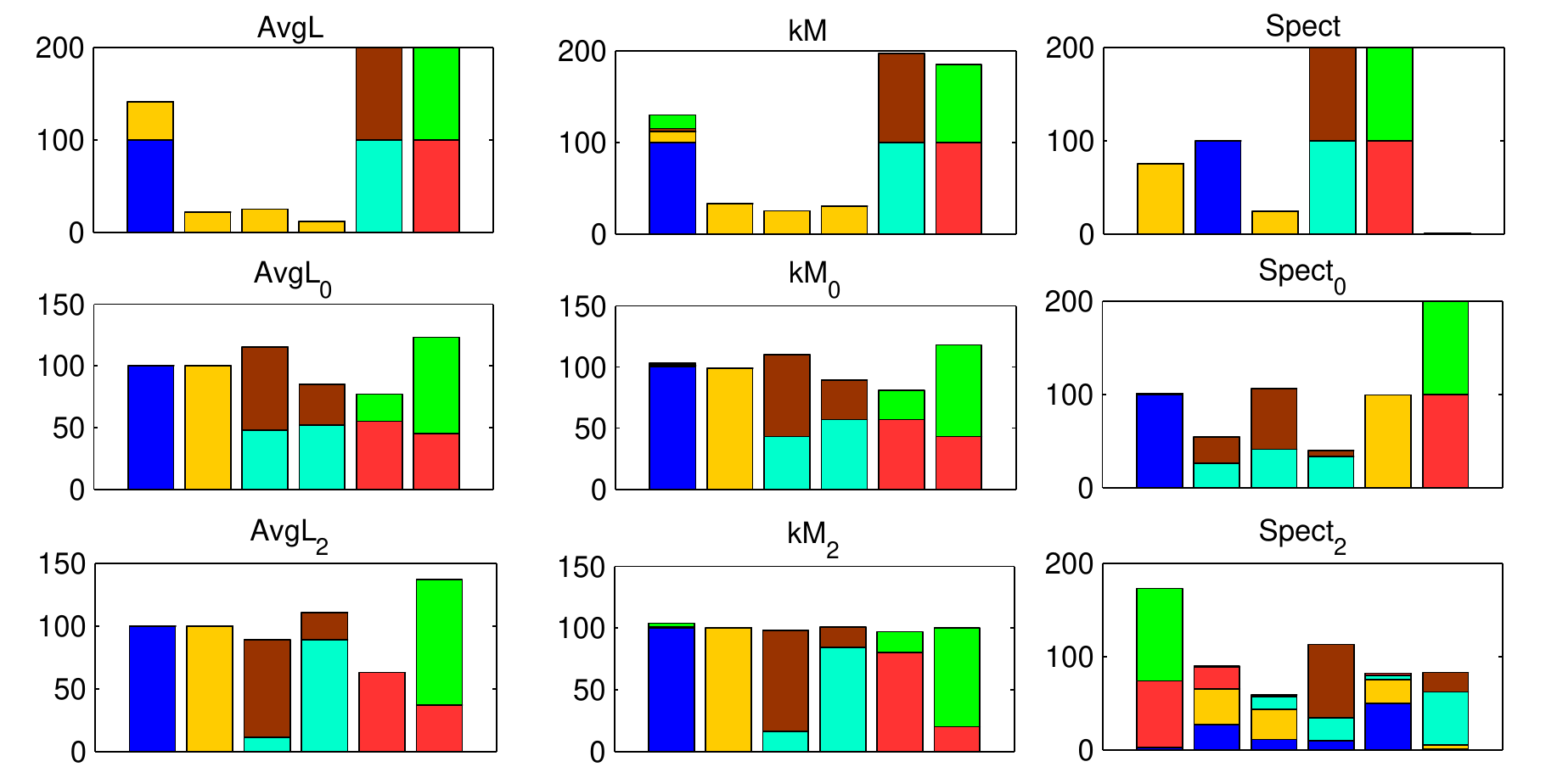}
\end{subfigure}
\vspace{-0.5in}
\caption{\small Compositions of (a) four and (b) six estimated clusters. Each bar corresponds to a single cluster, comprising of observations from different classes (normal(\crule[normal]{0.1in}{0.1in}), cyclic(\crule[cyclic]{0.1in}{0.1in}), increasing trend(\crule[inc]{0.1in}{0.1in}), decreasing trend(\crule[dec]{0.1in}{0.1in}), upward shift(\crule[up]{0.1in}{0.1in}) and downward shift(\crule[down]{0.1in}{0.1in})).}
\label{fig::synthetic_cluster_k=4.6}
\vspace{-0.1in}
\end{figure}

We also divided the data set into six clusters. In that situation, clustering algorithms based on MADD (both $\rho_0$ and $\rho_2$) performed better than their Euclidean counterparts (see Figure~\ref{fig::synthetic_cluster_k=4.6}(b)). For instance, while AvgL and kM put many of the {\tt N} and {\tt C} samples in the same cluster, AvgL$_0$, kM$_0$, AvgL$_2$ and kM$_2$ successfully separated normal ({\tt N}), cyclic ({\tt C}), upward ({\tt IT}, {\tt US}) and downward ({\tt DT}, {\tt DS}) patterns. However, none of them could completely distinguish between {\tt IT} and {\tt US} samples or {\tt DT} and {\tt DS} samples. This is quite expected from the plot of the observations in Figure~\ref{fig::Synthetic}. Spectral clustering algorithms failed to have satisfactory performance in this case.

\vspace{-0.250in}
\section{Concluding remarks}\label{sec::remarks}

\vspace{-0.075in}
In high dimensions, concentration of Euclidean distance often leads to poor performance by clustering algorithms based on it. In this article, we have used a data driven dissimilarity measure, called MADD, which takes care of this problem. Clustering algorithms based on MADD can lead to perfect clustering for HDLSS data even when those based on Euclidean distance perform miserably. We have amply demonstrated it in this article using theoretical as well as numerical results. 
While MDP clustering performs poorly for  HDLSS data with populations not differing in their locations, MADD based clustering algorithms can have excellent performance even when the populations have the same location and scale. Using suitable transformation $\psi$ on each covariate, MADD is able to distinguish between populations with different marginal distributions. However, instead of applying $\psi$ on each co-ordinate, one can divide $\Xvec$ into  disjoint blocks $\Xvec = (\Xvec^{(1)},\ldots,\Xvec^{(d_0)})^\top$ and define $\varphi_{h,\psi}(\Xvec,\Zvec)=h\big\{\sum_{q=1}^{d_0} \psi(\|\Xvec^{(q)}-\Zvec^{(q)}\|)\big\}$. MADD can be defined accordingly. If the sizes of these blocks are uniformly bounded, $\rho^\ast_{h,\psi}(i,j)$ turns out to be positive unless the $i$-th and the $j$-th populations have the same block distributions. Naturally one would like to have nearly independent blocks, but a suitable algorithm needs to be developed for this purpose.

For most of the data sets analyzed in this article, the spectral clustering algorithm of \cite{SM00} also worked better when a MADD based similarity measure was used. We observed the same for the spectral clustering algorithm of \cite{NJW02} as well, but to save space, we decided not to report them in this article.
Throughout this article, we have used AvgL for hierarchical clustering. However, other linkage methods like single linkage, complete linkage, Ward's linkage or centroid linkage \citep[see, e.g.,][]{DHS12,JW14} can also be used. One can prove the perfect clustering property of MADD versions of these linkage algorithms following the same line of arguments as used in the proofs of Theorems~\ref{thm::exact cluster hierarchical}, \ref{thm::MADD_separation_CLT} and \ref{thm::MADD_Euclid_CLT}.

We have also considered the problem of estimating the number of clusters and seen that the methods based on \textsc{Jump} statistic and \textsc{KL} statistic usually perform better in high dimensions when their MADD versions are used. We have also successfully used MADD versions of {\sc Dunn} index and penalized {\sc Dunn} index for this purpose. Under appropriate regularity conditions, the methods based on {\sc KL} statistic, {\sc Dunn} index and penalized {\sc Dunn} index turn out to be consistent in HDLSS asymptotic regime when AvgL($h,\psi$) or kM($h, \psi$) is used for base clustering. But, the choice of penalty function $\zeta$ in the penalized {\sc Dunn} index still remains an issue to be resolved. Throughout this article, we have used $\zeta(d)=\lambda \log(d)$, which was chosen based on our empirical experience. But, a suitable data driven choice of $\zeta$ may further improve the empirical performance of different clustering algorithms.

\vspace{-0.25in}
\section*{Acknowledgment}
\vspace{-0.1in}
We are grateful to Dr. J. Ahn for providing us with the codes for MDP clustering.

\vspace{-0.25in}
\section*{Appendix: Proofs and mathematical details}

\vspace{-0.075in}
{\bf Proof of Lemma~\ref{lemma::dist-convergence}:} Since $\Xvec \sim {\cal U}_d(a_1,b_1)$, the distribution function of $R = d^{-1/2}\|\Xvec\|$ is given by $F_R(r)=(r^d-a_1^d)/(b_1^d-a_1^d)$ for $a_1 \le r \le b_1$, $0$ for $r<a_1$ and $1$ for $r>b_1$. So, $R$ has the density $f_R(r) = dr^{d-1}/(b_1^d-a_1^d)$ for $a_1 \le r \le b_1$ and $0$ otherwise. Therefore, 
\begin{align}\label{eqn::ExB_1}
& E(d^{-1} \|\Xvec\|^2) = E(R^2) 
= \frac{d}{d+2} \frac{b_1^{d+2}-a_1^{d+2}}{b_1^d-a_1^d} \rightarrow b_1^2, \text{ and } \nonumber\\
& E(d^{-2} \|\Xvec\|^4) = E(R^4) = 
\frac{d}{d+4} \frac{b_1^{d+4}-a_1^{d+4}}{b_1^d-a_1^d} \rightarrow b_1^4
\vspace{-0.15in}
\end{align}
as $d \rightarrow \infty$. This implies $Var(d^{-1}\|\Xvec\|^2) \rightarrow 0$ and hence $d^{-1} \|\Xvec\|^2 \stackrel{P}{\rightarrow} b_1^2$ as $d \to \infty$. Similarly, we have $d^{-1} \|\Yvec\|^2 \overset{P}{\rightarrow} b_2^2$ as $d \to \infty$.

\noindent
Now, it is enough to show that $d^{-1} \left\langle \Xvec, \Yvec \right\rangle \stackrel{P}{\rightarrow}0$ as $d \rightarrow \infty$. Here $\Xvec$ and $\Yvec$ are independent, and they are spherically symmetric about ${\bf 0}$ \citep[see][]{F90}. So, we have $E(d^{-1}\left\langle \Xvec, \Yvec \right\rangle) = d^{-1} \sum_{q=1}^d E(X^{(q)})E(Y^{(q)}) = 0$, and $E(X^{(q)}X^{(q^{\prime})}) = E(Y^{(q)}Y^{(q^{\prime})}) = 0$ for all $q \neq q^\prime$. Therefore, $Var (d^{-1} \left\langle \Xvec,\Yvec \right\rangle ) =E(d^{-2}\left\langle \Xvec, \Yvec \right\rangle^2)=$ $d^{-1} E({X^{(1)}}^2)E({Y^{(1)}}^2)=d^{-1}E(d^{-1}\|\Xvec\|^2)E(d^{-1}\|\Yvec\|^2)$. We have proved that $E(d^{-1}\|\Xvec\|^2) \rightarrow b_1^2$ and $E(d^{-1}\|\Yvec\|^2)\rightarrow b_2^2$ as $d \rightarrow \infty$. So, $Var (d^{-1} \left\langle \Xvec,\Yvec \right\rangle ) \rightarrow 0$ and hence $d^{-1} \left\langle \Xvec, \Yvec \right\rangle \stackrel{P}{\rightarrow}0$ as $d \rightarrow \infty$. \qed

\noindent
{\bf Proof of Lemma~\ref{lemma::semi-metric}}: Non-negativity of $\rho_{h,\psi}$ is obvious and symmetry comes from the fact that $\varphi_{h,\psi}$ is symmetric. When $n=3$, we get
\vspace{-0.05in}
\begin{align*}
|\varphi_{h,\psi}(\xvec_1,\xvec_3) - \varphi_{h,\psi}(\xvec_2,\xvec_3)| &= |\varphi_{h,\psi}(\xvec_1,\xvec_3) - \varphi_{h,\psi}(\xvec_1,\xvec_2) + \varphi_{h,\psi}(\xvec_1,\xvec_2) - \varphi_{h,\psi}(\xvec_2,\xvec_3)| \nonumber \\
\vspace{-0.05in}
& \le |\varphi_{h,\psi}(\xvec_1,\xvec_2) - \varphi_{h,\psi}(\xvec_3,\xvec_2)| + |\varphi_{h,\psi}(\xvec_2,\xvec_1) - \varphi_{h,\psi}(\xvec_3,\xvec_1)|
\vspace{-0.4in}
\end{align*}
When $n \ge 4$, for any $k=4,\ldots,n$, we get
\vspace{-0.05in}
\begin{align*}
|\varphi_{h,\psi}(\xvec_1,\xvec_k) - \varphi_{h,\psi}(\xvec_2,\xvec_k)| &= |\varphi_{h,\psi}(\xvec_1,\xvec_k) - \varphi_{h,\psi}(\xvec_3,\xvec_k) + \varphi_{h,\psi}(\xvec_3,\xvec_k) - \varphi_{h,\psi}(\xvec_2,\xvec_k)| \nonumber \\
\vspace{-0.05in}
& \le |\varphi_{h,\psi}(\xvec_1,\xvec_k) - \varphi_{h,\psi}(\xvec_3,\xvec_k)| + |\varphi_{h,\psi}(\xvec_2,\xvec_k) - \varphi_{h,\psi}(\xvec_3,\xvec_k)|
\vspace{-0.5in}
\end{align*}
Combining these two facts, we have
\vspace{-0.1in}
\begin{align*}
&\sum_{k \ne 1,2} |\varphi_{h,\psi}(\xvec_1,\xvec_k) - \varphi_{h,\psi}(\xvec_2,\xvec_k)|\\
&~~~ \le \sum_{k \ne 1,3} |\varphi_{h,\psi}(\xvec_1,\xvec_k) - \varphi_{h,\psi}(\xvec_3,\xvec_k)| + \sum_{k \ne 2,3} |\varphi_{h,\psi}(\xvec_2,\xvec_k) - \varphi_{h,\psi}(\xvec_3,\xvec_k)|~~~~~~~~~ \qed
\end{align*}

\vspace{0.05in}
\noindent
{\bf Proof of Lemma~\ref{lemma:MADD_asymptotic}:} The proof follows from our discussion preceding the statement of the lemma. Hence it is omitted.

\vspace{0.05in}
\noindent
{\bf Proof of Lemma~\ref{lemma:MADD_generalasymptotic}:} If the $i$-th and the $j$-th populations have the same marginal distributions, then $\varphi_{h,\psi}^\ast(i,\ell) = \varphi_{h,\psi}^\ast(j,\ell)$ for all $\ell=1,\ldots,k_0$. As a result, we have $\rho^\ast_{h,\psi}(i,j) = 0$.

For the only if part, first observe that $\rho^\ast_{h,\psi}(i,j) \ge (n_i-1) \bigl|\varphi_{h,\psi}^\ast(i,j) - \varphi_{h,\psi}^\ast(i,i) \bigr| + (n_j - 1) \bigl|\varphi_{h,\psi}^\ast(i,j) - \varphi_{h,\psi}^\ast(j,j) \bigr|$. Now, if the right side is zero, we have $\varphi_{h,\psi}^\ast(i,j)=\varphi_{h,\psi}^\ast(i,i)$ and $\varphi_{h,\psi}^\ast(i,j)=\varphi_{h,\psi}^\ast(j,j)$. Since $h$ is a one-to-one function, this implies $d^{-1} \sum_{q=1}^d \bigl\{2E \psi(|X_1^{(q)}-Y_1^{(q)}|) - E \psi(|X_1^{(q)}-X_2^{(q)}|) - E \psi(|Y_1^{(q)}-Y_2^{(q)}|) \bigr\} = 0$, where $\Xvec_1,\Xvec_2$ and $\Yvec_1,\Yvec_2$ are independent observations from the $i$-th and the $j$-th populations, respectively. Now, since $\psi^\prime(t)/t$ is strictly monotone, each summand in the left side is positive, and it is zero if and only if the respective marginal distributions are equal (see \cite{BF10,BMG15}). Thus, $\rho^\ast_{h,\psi}(i,j) = 0$ implies that the $i$-th and the $j$-th populations have the same marginals. \qed

\noindent
{\bf Proof of Lemma~\ref{lemma:MADD_euclidasymptotic}:} If $\rho_{h,\psi}^{\ast}(i,j)=0$, then for $\ell=1,\ldots,k_0$, we have $\varphi_{h,\psi}^{\ast}(i,\ell) = \varphi_{h,\psi}^{\ast}(j,\ell)$. Now, for $h(t)=\sqrt{t}$ and $\psi(t)=t^2$, we have $\varphi_{h,\psi}^{\ast}(i,\ell) = d^{-1/2} \sqrt{tr(\sigmat_i) + tr(\sigmat_\ell) + \|\muvec_i-\muvec_\ell\|^2}$ and $\varphi_{h,\psi}^{\ast}(j,\ell) = d^{-1/2} \sqrt{tr(\sigmat_j) + tr(\sigmat_\ell) + \|\muvec_j-\muvec_\ell\|^2}$. So, $\rho_{h,\psi}^{\ast}(i,j)=0$ if and only if $\|\muvec_i-\muvec_\ell\|^2+tr(\sigmat_i) = \|\muvec_j - \muvec_\ell\|^2+ tr(\sigmat_j)$ for every $\ell=1,\ldots,k_0$. Therefore, taking $\ell=i$ and $\ell=j$, we get $tr(\sigmat_i) = tr(\sigmat_j)$ and $\|\muvec_i -\muvec_j\|=0$. On the other hand, if $tr(\sigmat_i) = tr(\sigmat_j)$ and $\muvec_i = \muvec_j$, it is easy to check that $\rho_{h,\psi}^{\ast}(i,j) = 0$. \qed

\vspace{0.05in}
\noindent
{\bf Proof of Theorem~\ref{thm::exact cluster hierarchical}:} From Lemma~\ref{lemma:MADD_asymptotic} and $(A2)$, we have $\rho_{h,\psi}(\Xvec,\Yvec) \stackrel{P}{\rightarrow} 0$ when $\Xvec,\Yvec$ come from the same population, but when they are from different populations, we have $\rho_{h,\psi}(\Xvec,\Yvec) > 0$ for all but finitely many $d$. So, for every $k$ and $i \ne j$, we get
\vspace{-0.05in}
\begin{equation}
\label{eq::exact cluster hierarchical-1}
\Pr\Bigl(\max_{\Xvec,\Yvec \in {\cal X}_{k}} \rho_{h,\psi}(\Xvec,\Yvec) < \min_{\Xvec \in {\cal X}_i, \Yvec \in {\cal X}_j} \rho_{h,\psi}(\Xvec,\Yvec)\Bigr) \to 1 \text{ as } d \to \infty.
\vspace{-0.05in}
\end{equation}
Therefore, at the first step of AvgL($h,\psi$), two members of the same population merge together with probability converging to $1$ as $d \to \infty$. Now at any step $r$ $(2 \le r < n-k_0)$, given that observations from the same population were merged together at each of the ($r-1$) previous steps, any cluster $C$ becomes a subset of ${\cal X}_k$ for some $k$, and we have
\vspace{-0.05in}
\begin{equation}
\Pr \Bigl( \max_{k}\max_{C,C^\prime \subset {\cal X}_{k}} \Delta(C,C^\prime) < \min_{i \neq j} \min_{C \subset {\cal X}_{i}, C^\prime \subset {\cal X}_{j}} \Delta(C,C^\prime) \Bigr) \rightarrow 1 \text{ as } d \to \infty,
\vspace{-0.05in}
\end{equation}
where $\Delta(C,C^\prime) = {(|C||C^\prime|)}^{-1} \sum_{\Xvec \in C, \Yvec \in C^\prime} \rho_{h,\psi}(\Xvec,\Yvec)$. Therefore, two clusters containing observations from the same population will merge with probability tending to $1$ as $d \to \infty$. Since $k_0$ is known, these two facts together prove the result.\qed

\noindent
{\bf Proof of Theorem~\ref{thm::exact cluster kmeans}:} Note that for any $k$, $|C_k|^{-1} \sum_{\Zvec \in C_k, \Zvec \ne \Xvec} \rho_{h,\psi}^2(\Xvec,\Zvec) \overset{P}{\rightarrow} 0$ as $d \to \infty$ if and only if $\Xvec$ and all observations in $C_k$ are from the same population (follows from the proof of Theorem~\ref{thm::exact cluster hierarchical}). So, if each $C_k$ ($k=1,\ldots,k_0$) contains observations from the same population, $\Phi^\ast(C_1,\ldots,C_{k_0}) \stackrel{P}{\rightarrow} 0$ as $d \to \infty$. Otherwise, we have $\liminf_{d \to \infty} \Phi^\ast(C_1,\ldots,C_{k_0}) > 0$ (follows from $(A2)$). So, when $k_0$ is known, for the minimization of $\Phi^\ast(C_1,\ldots,C_{k_0})$, each $C_k$ must contain all observations from a single population with probability converging to one as the dimension increases. This proves the convergence of the Rand index to zero.\qed

\noindent
{\bf Proof of Theorem~\ref{thm::MADD_separation_CLT}:}
Let $\Xvec$ and $\Zvec$ be independent observations from $i$-th and $\ell$-th populations $(i,\ell=1,\ldots,k_0)$, and define $V_{d} = d^{-1}\sum_{q=1}^d \psi(|X^{(q)}-Z^{(q)}|)$. Since $\big(V_{d}-E(V_{d})\big)/\sqrt{Var(V_{d})} = {\bf O}_P(1)$, we have $V_{d} - E(V_{d}) = {\bf O}_P(\vartheta(d)/d)$. Since $h$ is Lipschitz continuous, this implies $\bigl|\varphi_{h,\psi}(\Xvec,\Zvec) - \varphi_{h,\psi}^\ast(i,\ell)\bigr| = |h(V_d) - h(E(V_d))| \le C_0 |V_d - E(V_d)| = {\bf O}_P(\vartheta(d)/d)$. So, for an independent observation $\Yvec$ from the $j$-th population, we get $\bigl|\varphi_{h,\psi}(\Xvec,\Zvec) - \varphi_{h,\psi}(\Yvec,\Zvec)\bigr| = \bigl|\varphi_{h,\psi}^{\ast}(i,\ell) - \varphi_{h,\psi}^{\ast}(j,\ell)\bigr| + {\bf O}_P(\vartheta(d)/d)$ as $d \to \infty$. Since the number of observations is finite, we get $\rho_{h,\psi}(\Xvec,\Yvec) = \rho_{h,\psi}^{\ast}(i,j) + {\bf O}_P(\vartheta(d)/d)$. Now, for all $i=1,\ldots,k_0$, $\rho_{h,\psi}^{\ast}(i,i) = 0$, while for all $i \neq j$, $\rho_{h,\psi}^{\ast}(i,j)$ has asymptotic order higher than that of $\vartheta(d)/d$. Therefore, for $\Xvec,\Yvec$ from the same population and $\Xvec^\prime,\Yvec^\prime$ from different populations we get $\Pr\big(\rho_{h,\psi}(\Xvec,\Yvec) < \rho_{h,\psi}(\Xvec^\prime,\Yvec^\prime)\big) \to 1$ as $d \to \infty$. Now, the proof follows using the same line of arguments as used in the proofs Theorems \ref{thm::exact cluster hierarchical} and \ref{thm::exact cluster kmeans}. \qed

\noindent
{\bf Proof of Theorem~\ref{thm::MADD_Euclid_CLT}:} Let $\Xvec$ and $\Zvec$ be two independent observations from $i$-th and $\ell$-th populations $(i,\ell=1,\ldots,k_0)$. Note that for $\rho_0$, we use $h(t)=\sqrt{t}$ and $\psi(t) = t^2$. Therefore, taking $V_d = d^{-1}\sum_{q=1}^d (X^{(q)}-Z^{(q)})^2$, we get $\varphi_{h,\psi}(\Xvec,\Zvec) - \varphi_{h,\psi}^{\ast}(i,\ell) = \sqrt{V_d} - \sqrt{E(V_d)} = (V_d-E(V_d))/\big(\sqrt{V_d} + \sqrt{E(V_d)}\big)$, where $E(V_d) = d^{-1}\big\{\|\muvec_i-\muvec_{\ell}\|^2 + tr(\sigmat_i+\sigmat_{\ell})\big\} \ge d^{-1}tr(\sigmat_i)$. So, $\sqrt{dE(V_d)/\vartheta(d)}$ remains bounded away from $0$, and hence $\sqrt{\vartheta(d)}/\big(\sqrt{d V_d}+\sqrt{d E(V_d)}\big)$ remains bounded as $d \to \infty$. Now, $(V_d-E(V_d))/\sqrt{Var(V_d)}={\bf O}_p(1)$ implies $(V_d-E(V_d))={\bf O}_p(\vartheta(d)/d)$. Again, $1/\big(\sqrt{V_d} + \sqrt{E(V_d)}\big) = {\bf O}_p(\sqrt{d/\vartheta(d)})$. So, $\varphi_{h,\psi}(\Xvec,\Zvec) = \varphi_{h,\psi}^{\ast}(i,\ell) + {\bf O}_{P}(\sqrt{\vartheta(d)/d})$, and hence we have $\rho_0(\Xvec,\Yvec) = \rho_0^{\ast}(i,j) + {\bf O}_{P}(\sqrt{\vartheta(d)/d})$. So, following the proof of Theorem \ref{thm::MADD_separation_CLT}, one can show that AvgL$_0$ and kM$_0$ will have the perfect clustering property if for every $i \neq j$, $\sqrt{d}\rho_0^\ast(i,j)/\sqrt{\vartheta(d)} \to \infty$ or $ d \rho_0^{\ast 2}(i,j)/{\vartheta(d)} \to \infty$ as $d \to \infty$.
Now, from the proof of Lemma~\ref{lemma:MADD_euclidasymptotic}, it follows that if $\|\muvec_i-\muvec_j\|^2/\vartheta(d) \to \infty$ and/or $|tr(\sigmat_i)-tr(\sigmat_j)|/\vartheta(d) \to \infty$, then $ d \rho_0^{\ast 2}(i,j)/{\vartheta(d)} \to \infty$ as $d \to \infty$. \qed

All estimation methods that we discuss henceforth are based on $\rho_{h,\psi}$. So, we have $W_k = \sum_{j=1}^k {(2|C_j|)}^{-1} \sum_{\zvec,\wvec \in C_j} \rho_{h,\psi}^2(\zvec,\wvec)$, $\Delta_0(C_i)={\{|C_i|(|C_i|-1)\}}^{-1}\sum_{\zvec,\wvec \in C_i} \rho_{h,\psi}(\zvec,\wvec)$, and $\Delta(C_i,C_j) = {(|C_i||C_j|)}^{-1} \sum_{\zvec \in C_i, \wvec \in C_j} \rho_{h,\psi}(\zvec,\wvec)$.

\noindent
{\bf Proof of Theorem~\ref{thm::consistency Dunn}:} $(i)$ Since the base clustering algorithm is POP at $k_0$, for any $k<k_0$, there exists at least one estimated cluster which contains observations from two different populations, and no two clusters contain observations from the same population. So, under $(A3)$, we have $\Delta_0(C_i) \overset{P}{\asymp} \phi_{+}(d)$ for some $i$ and $\Delta(C_i,C_j) \overset{P}{\asymp} \phi_{+}(d)$ for every $i \ne j$. Thus, $B_k^\circ = \min_{1 \le i < j \le k} \Delta(C_i,C_j) \overset{P}{\asymp} \phi_{+}(d)$ and $W_k^\circ = \max_{1 \le i \le k} \Delta_0(C_i) \overset{P}{\asymp} \phi_{+}(d)$, and hence we get $\textsc{D}(k) =B_k^\circ/W_k^\circ \overset{P}{\asymp} 1$.

\noindent
For $k>k_0$, no cluster contains observations from two different populations, while there exists at least two clusters which contain observations from the same population. So, $\Delta_0(C_i) \overset{P}{\asymp} \phi_{-}(d)$ for every $i$ and $\Delta(C_i,C_j) \overset{P}{\asymp} \phi_{-}(d)$ for some $i \ne j$. Thus, $B_k^\circ \overset{P}{\asymp} \phi_{-}(d)$, $W_k^\circ \overset{P}{\asymp} \phi_{-}(d)$, and hence $\textsc{D}(k) \overset{P}{\asymp} 1$.

\noindent
For $k=k_0$, each cluster contains observations from same population and two different clusters contain observations from two different populations. This implies that $\Delta_0(C_i) \overset{P}{\asymp} \phi_{-}(d)$ for every $i$ and $\Delta(C_i,C_j) \overset{P}{\asymp} \phi_{+}(d)$ for every $i \ne j$. So, we have $B_k^\circ \overset{P}{\asymp} \phi_{+}(d)$ and $W_k^\circ \overset{P}{\asymp} \phi_{-}(d)$ and hence $\textsc{D}(k) \overset{P}{\asymp} \big(\phi_{+}(d)/\phi_{-}(d)\big)$.

\noindent
Combining these three cases, and noting that $\phi_{-}(d) = {\bf o}\big(\phi_{+}(d)\big)$, we get $\Pr\big(\textsc{D}(k_0) > \textsc{D}(k)~\forall k \ne k_0\big) \to 1$ as $d \to \infty$. This implies $\hat{k}^\ast_D \stackrel{P}{\rightarrow} k_0$ as $d \to \infty$.

\noindent
$(ii)$ For all $k\ge 2$, the \textsc{KL} statistic can be written as $\textsc{KL}(k)$=$\big|(\Lambda_{k-1}-\Lambda_k)/(\Lambda_{k}-\Lambda_{k+1})\big|$, where $\Lambda_k=k^{2/d}W_k$. Since the base clustering algorithm is POP at $k_0$, from our discussion in part $(i)$, it follows that $W_k \overset{P}{\asymp} \phi_{+}(d)$ for $k < k_0$ and  $W_k \overset{P}{\asymp} \phi_{-}(d)$ for $k \ge k_0$. Note that, for any fixed $k$, $k^{{2}/{d}} \to 1$ as $d \to \infty$. So, for all $k \ge 1$, $\Lambda_k \overset{P}{\asymp} W_k$. Now, it is easy to check that $\textsc{KL}(k_0) \overset{P}{\asymp} \big(\phi_{+}(d)/\phi_{-}(d)\big)$ and $\textsc{KL}(k) \overset{P}{\asymp} 1$ for all other values of $k$. Therefore, $\Pr\big(\textsc{KL}(k_0) > \textsc{KL}(k)~~\forall k \ne k_0\big) \to 1$ as $d \to \infty$, and hence we have ${\hat k}^\ast_{KL} \overset{P}{\rightarrow} k_0$ as $d \to \infty$.

\noindent
$(iii)$ Note that $\hat{d}_k = W_k \overset{P}{\asymp} \phi_{+}(d)$ and $\phi_{-}(d)$ for $1 \le k < k_0$ and $k \ge k_0$, respectively (follows from our discussion in parts $(i)$ and $(ii)$). So, we have $\textsc{Jump}(k) = \hat{d}^{-1}_k - \hat{d}^{-1}_{k-1} \overset{P}{\asymp} 1/\phi_{+}(d)$ or $1/\phi_{-}(d)$ according as $k < k_0$ or $k \ge k_0$. This implies that $\Pr\big({\hat k}^\ast_J < k_0\big) \to 0$ as $d \to \infty$. \qed

\noindent
{\bf Proof of Theorem~\ref{thm::consistency penalized Dunn}:} First consider the case $k_0 \ge 2$. While proving part $(i)$ of Theorem~\ref{thm::consistency Dunn}, we have shown that in this case, $\textsc{D}(k_0) \overset{P}{\asymp} \big(\phi_{+}(d)/\phi_{-}(d)\big)$ and $\textsc{D}(k) \overset{P}{\asymp} 1$ for all other choice of $k \ge 2$. Also, we have $W_1^\circ \overset{P}{\asymp} \phi_{+}(d)$ and $B_1^\circ \overset{def}{=} B_2^\circ \overset{P}{\asymp} \phi_{+}(d)$, which implies $\textsc{D}(1)\overset{def}{=} B_1^\circ / W_1^\circ \overset{P}{\asymp} 1$. So, for any $k \neq k_0$, $\textsc{PD}(k_0) -\textsc{PD}(k) = \textsc{D}(k_0) - \textsc{D}(k) - (k_0-k) \zeta(d) \overset{P}{\rightarrow} \infty$ as $d \rightarrow \infty$ (since $\zeta(d) = {\bf o}\big(\phi_{+}(d)/\phi_{-}(d)\big)$). Thus, ${\hat k}_{PD}^* \stackrel{P}{\rightarrow} k_0$ as $d \rightarrow \infty$.

\noindent
When $k_0 = 1$, we have $W_k^\circ \overset{P}{\asymp} \phi_{-}(d)$, $B_k^\circ \overset{P}{\asymp} \phi_{-}(d)$ for every $k \ge 1$. So, $\textsc{PD}(1) - \textsc{PD}(k) = \textsc{D}(1) - \textsc{D}(k) +(k-1) \zeta(d) \overset{P}{\rightarrow} \infty \text{ as } d \to \infty$ (since $\textsc{D}(k) \overset{P}{\asymp} 1$ for $k \ge 1$ and $\zeta(d) \to \infty$ as $d \to \infty$) and hence ${\hat k}_{PD}^* \stackrel{P}{\rightarrow} k_0$.\qed

\vspace{-0.1in}
\begin{lemma}\label{lemma:MADD_sameclass}
Let $\Xvec$ and $\Yvec$ be two independent observations from the $i$-th population. For an independent observation $\Zvec$ from the $j$-th population $(j=1,\ldots,k_0)$, assume that $Var\big\{\sum_{q=1}^d \psi(|X^{(q)}-Z^{(q)}|) \big\} = {\bf O}(d^{2-\epsilon_0})$ for some $\epsilon_0 > 0$. If $h$ is H{\"o}lder continuous with exponent $\gamma$, then $\rho_{h,\psi}(\Xvec,\Yvec) = {\bf O}_P(d^{-\alpha_0})$ with $\alpha_0 = \gamma\epsilon_0/2$.
\vspace{-0.05in}
\end{lemma}
\noindent
{\bf Proof:} Define $V_d = d^{-1} \sum_{q=1}^d \psi(|X^{(q)} - Z^{(q)}|)$ and $V_d^\prime = d^{-1} \sum_{q=1}^d \psi(|Y^{(q)} - Z^{(q)}|)$, for some $\Zvec \ne \Xvec,\Yvec$. Note that $V_d - V_d^\prime = (V_d - E V_d) - (V_d^\prime - E V_d^\prime)$. Now, write
$$V_d - E V_d = \frac{V_d - E V_d}{\sqrt{Var(V_d)}}  \sqrt{Var(V_d)}.$$
The first term on the right side is ${\bf O}_P(1)$, and under the given condition, the second term is ${\bf O}(d^{-\epsilon_0/2})$. So, we have $V_d - E V_d = {\bf O}_P(d^{-\epsilon_0/2})$. Similarly, one gets $V_d^\prime - E V_d^\prime = {\bf O}_P(d^{-\epsilon_0/2})$. Thus, $V_d - V_d^\prime = {\bf O}_P(d^{-\epsilon_0/2})$. Now, since $h$ is H{\"o}lder continuous with exponent $\gamma$, we get $\bigl|\varphi_{h,\psi}(\Xvec,\Zvec) - \varphi_{h,\psi}(\Yvec,\Zvec)\bigr| =\bigl|h(V_d) - h(V_d^\prime)\bigr| \le C_0\bigl|V_d - V_d^\prime\bigr|^\gamma = {\bf O}_P(d^{-\gamma \epsilon_0/2})={\bf O}_P(d^{-\alpha_0})$.
Since $n$ is finite, this in turn proves that $\sum_{\Zvec \ne \Xvec,\Yvec} |\varphi_{h,\psi}(\Xvec,\Zvec) - \varphi_{h,\psi}(\Yvec,\Zvec)| = {\bf O}_P(d^{-\alpha_0})$.\qed

\begin{remark}
{\rm For $\rho_0$ (i.e., $h(t) = \sqrt{t}$ and $\psi(t) = t^2$), $\varphi_{h,\psi}(\Xvec,\Zvec)- \varphi_{h,\psi}(\Yvec,\Zvec) = h(V_d) - h(V_d^\prime) = (V_d - V_d^\prime) (2\sqrt{\xi_d})^{-1}$, where $\xi_d$ lies between $V_d$ and $V_d^\prime$. Also, $V_d = d^{-1} \|\Xvec - \Zvec\|^2$ remains bounded away from $0$ in probability (and so does $V_d^\prime$). Therefore, $\xi_d^{-1/2} = {\bf O}_P(1)$, and hence $\varphi_{h,\psi}(\Xvec,\Zvec) - \varphi_{h,\psi}(\Yvec,\Zvec) = {\bf O}_P(d^{-\epsilon_0/2})$. So, the H{\"o}lder continuity of $h$ is only sufficient, but not necessary.}
\end{remark}

\setlength{\parindent}{0pt}
\spacingset{1}
\setlength{\bibsep}{0.1in}
\small
\bibliographystyle{rss}
\bibliography{CLUSTER}

\end{document}